# Towards Developing Mid-Infrared Photonics Using Mxenes

**Yas Al-Hadeethi**[(1),(3),(4)†]**, Chandraman Patil**[(2)†]**, Wafa Said Bait Haridh**[(1)]**, Moustafa Ahmed**[(1)]**, Jamaan E. Alassafi**[(1)]**, Nada M. Bedaiwi**[(1)]**, Elham Heidari**[(2)]**, Hamed Dalir**[(2)*]

[1] Physics Department, Faculty of Science, King Abdulaziz University, Jeddah 21589, Saudi Arabia
[2] Department of Electrical & Computer Engineering, University of Florida, Gainesville, FL 32611, USA
[3] Lithography in Devices Fabrication and Development Research Group, Deanship of Scientific Research, King Abdulaziz University, Jeddah 21589, Saudi Arabia
[4] King Fahd Medical Research Center (KFMRC), King Abdulaziz University, Jeddah 21589, Saudi Arabia

*Corresponding Author: hamed.dalir@ufl.edu
†Authors contributed equally

## Abstract

Recent research and development in the mid-infrared (IR) wavelength range (2-20 um) for a variety of applications, such as trace gas monitoring, thermal imaging, and free space communications have shown tremendous and fascinating progress. MXenes, which mainly refer to two-dimensional (2D) transition-metal carbides, nitrides, and carbonitrides, have drawn a lot of interest since their first investigation in 2011. MXenes project enormous potential for use in optoelectronics, photonics, catalysis, and energy harvesting fields proven by extensive experimental and theoretical studies over a decade. MXenes offers a novel 2D nano platform for cutting-edge optoelectronics devices due to their interesting mechanical, optical, and electrical capabilities, along with their elemental and chemical composition. We here discuss the key developments of MXene emphasizing the evolution of material synthesis methods over time and the resulting device applications. Photonic and optoelectronic device design and fabrication for mid-IR photonics are demonstrated by integrating MXene materials with various electrical and photonic platforms. Here, we show the potential of using Mxene in photonics for mid-IR applications and a pathway toward achieving next-generation devices for various applications.

## Introduction

The mid-infrared (mid-IR) spectral range holds significant importance in various scientific and technological applications owing to reduced complexity in platform development and integration. The unique molecular fingerprint of the mid-IR spectrum enables precise identification and characterization of chemical compounds, making it crucial for chemical sensing, environmental monitoring, biomedical diagnostics, and industrial process control **Fig. 1**. Additionally, mid-IR photonics play a vital role in defense and security applications, including infrared countermeasures and standoff detection of hazardous materials. Over the past few decades, noteworthy progress has been made in the field of mid-IR photonics, driven by advancements in materials, devices, and fabrication techniques. Various research efforts have focused on developing efficient mid-IR sources, enabling the generation of coherent and broadband light in this spectral range. Quantum cascade lasers (QCLs), based on inter-sub-band transitions in semiconductor heterostructures, have emerged as highly efficient and tunable sources in the mid-IR region [1]. Additionally, inter-band cascade lasers (ICLs) and optical parametric oscillators (OPOs) have demonstrated their potential for mid-IR emission [2], [3]. To guide and manipulate mid-IR light, waveguides and fibers tailored for this spectral range are crucial. Chalcogenide glasses, such as $As_2S_3$ and $GeSe_2$, have shown promise as mid-IR waveguide materials due to their high transparency and low propagation loss [4]. Silicon and germanium-based waveguides have also gained attention, offering compatibility with established silicon photonics platforms [5]. Additionally, specialty fibers, including photonic crystal fibers and hollow-core fibers, have been designed to achieve low-loss transmission and dispersion control in the mid-IR range [6], [7].

Mid-IR components and devices are essential for efficient manipulation and detection of mid-IR light. Modulators capable of controlling the intensity, phase, and polarization of mid-IR signals have been developed, leveraging electro-optic, acoustic-optic, and opto-mechanical effects [8], [9]. Detectors with high responsivity and sensitivity in the mid-IR have been realized, including mid-IR photodiodes, quantum well-infrared photodetectors (QWIPs), and superconducting detectors [10], [11]. Furthermore, beam steering and shaping devices, such as microelectromechanical systems (MEMS) mirrors and meta surfaces, enable precise control of mid-IR light propagation [12], [13]. Mid-IR spectrometers, both dispersive and Fourier-transform-based, offer high-resolution molecular spectroscopy in this spectral range [14], [15]. The development of mid-IR sensing and imaging systems has witnessed significant advancements in recent years. Mid-IR gas sensors utilizing absorption spectroscopy enable rapid and accurate detection of various gases, addressing critical applications such as environmental monitoring and industrial safety [16]. Chemical sensing techniques, including Raman spectroscopy and photoacoustic spectroscopy, offer label-free and sensitive analysis of chemical compounds in the mid-IR range [17]-[20]. Mid-IR imaging systems have found applications in non-destructive testing, thermal imaging, and biomedical imaging, providing valuable insights into material composition and biological structures [21]-[24]. As the field of mid-IR photonics continues to progress, several emerging trends and challenges shape its future. Integration and miniaturization of mid-IR photonic systems are of paramount importance to enable portable and point-of-care applications. The development of novel mid-IR materials, such as graphene, MXenes, and perovskites, holds promise for enhanced device performance and functionality [25]-[30]. Additionally, enhancing the efficiency, sensitivity, and dependability of mid-IR systems continues to be a critical issue that necessitates ongoing investigation by both researchers and engineers [31].

## General Application of MXenes

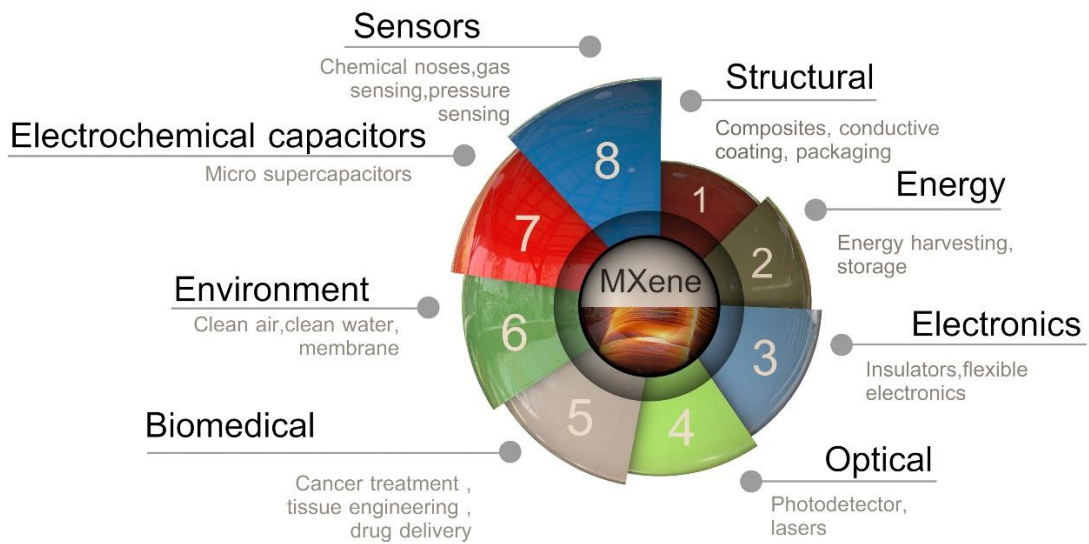

**Figure 1.** Fields of applications explored in research and development by integrating Mxene materials for the design and development of electrical, optical, mechanical, and chemical-based devices. Such materials provide hybrid, heterogeneous, and standalone platforms for developing next-generation devices.

Advances in mid-IR sources, waveguides, components, sensing, and imaging systems have significantly enhanced our capabilities in this spectral range. Continued research and development efforts are focused on addressing emerging trends and overcoming challenges, paving the way for further advancements in mid-IR photonics. 2D materials have attracted significant attention since the discovery of graphene, with their exceptional physical properties. In recent years, the integration of 2D layered materials has shown great potential for enhancing the functionalities of conventional devices [19], [20], [27]-[29], [32]-[34]. However, there is a lack of comprehensive reviews and understanding regarding the potential applications of MXenes in various research fields, including mass manufacturing, process integration, and general features. MXenes offers promising opportunities in diverse fields such as composites, energy, electronics, optics, biomedical applications, environmental monitoring, electrochemical capacitors, and sensors.

MXene is a rapidly growing family of 2D materials, characterized by n+1 (n = 1 to n=3) layers of early transition metals (M) interleaved with n layers of carbon or nitrogen (X) [35]-[37]. The general chemical formula of MXenes is $M_{n+1}X_nT_x$ (n = 1, 2, or 3), where M represents early d-block transition metals (e.g., Ta, Hf, Ti, Mo, Nb, Sc, Zr, Cr, V, and W), X represents carbon (C) and/or nitrogen (N), and Tx denotes surface groups such as ‾O, ‾OH, ‾Cl, and ‾F, which are bonded to the outer M layers [38], [39]. Atomic schematics of three types of MXenes are shown in Fig. 2. In conclusion, the exploration of MXenes and their potential applications have gained significant momentum. With their unique properties and versatile composition, MXenes hold great promise for advancements in various fields, including optoelectronics and photonics, contributing to the growing landscape of 2D materials research.

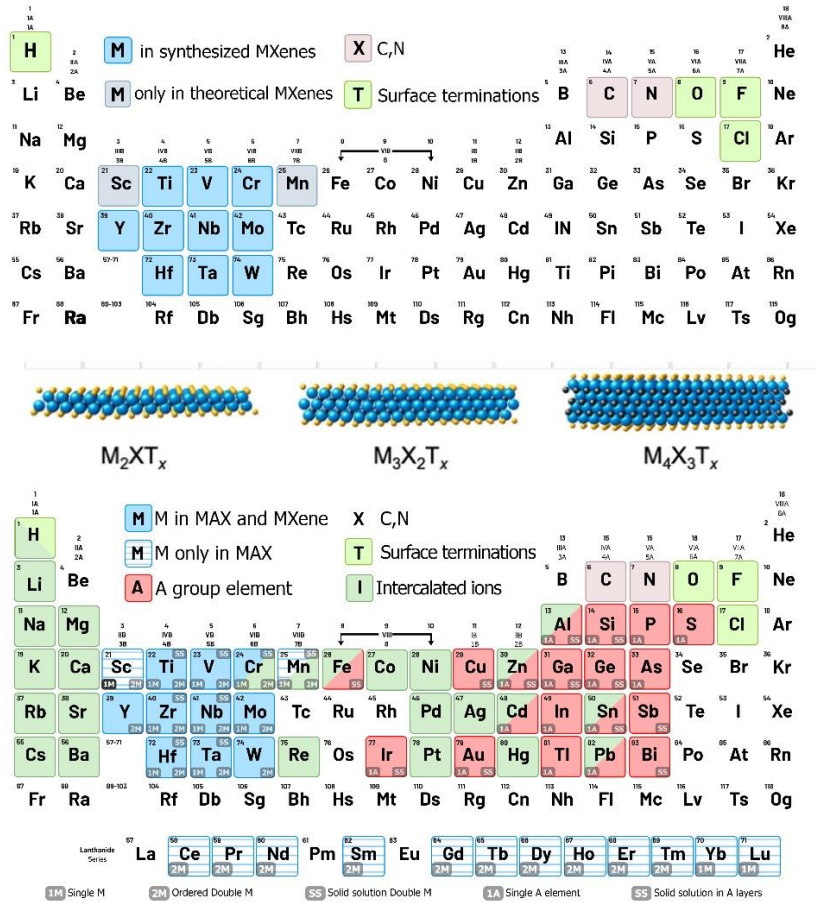

**Figure 2.** The periodic table highlights the compositions of MXenes and MAX phases. The numerous range of combinations of elements that can be used to grow the various materials in this class provide a wide range of integration.

Recent years have witnessed a significant surge in interest surrounding 2D materials that contain transition metal atoms, owing to their unique physical and chemical properties that distinguish them from their bulk counterparts. This has prompted the ongoing development of computational techniques, which explore a wide range of 2D materials and expand the potential applications of these materials. To anticipate future members of the MXene family and comprehend the electronic properties of MXenes, it is crucial to gain a theoretical understanding of their electronic characteristics and heterostructures. In this review article, we delve into the current advancements in theoretical research concerning MXenes and MXene-based heterostructures. We begin by examining the computational methods employed to generate theoretical insights into MXenes. Subsequently, we explore the anticipated topological features, and work functions, as well as the optical, magnetic, and electronic properties of MXenes as predicted through theoretical calculations. Finally, we provide a summary of the key attributes of heterostructures comprising MXenes and 2D layers.

The mid-infrared (IR) wavelength range spanning from 2-20 μm offers numerous applications, including chemical bond spectroscopy, trace gas sensing, and medical diagnostics. This wavelength range is particularly valuable due to the distinctive absorption spectra of many molecules within it, earning the designation of the "fingerprint region" for the 8-20 μm segment. Additionally, the mid-IR region, especially the atmospheric windows of 3-5 μm and 8-11 μm, is well-suited for thermal imaging and free-space communications. While significant progress has been made in the development of high-power, room-temperature mid-IR light sources, and sensitive detectors in the past decade or so, there has been limited effort in the advancement of passive photonic components for the mid-IR range. These components include waveguides, resonators, splitters, modulators, and more. Active optoelectronic devices such as lasers and detectors have traditionally employed III-IV materials, while chalcogenide glasses have been utilized in passive photonic elements for mid-IR photonics [40], [41].

However, the silicon photonics community has proposed a promising approach for the development of mid-IR complementary metal-oxide-semiconductor (CMOS)-compatible integrated photonic systems. This strategy aims to leverage the expertise and infrastructure of silicon photonics to enable the realization of integrated mid-IR photonic systems. The utilization of $Ti_3C_2T_x$ MXene in ultrafast photonics has shown significant promise, particularly in the mid-infrared (MIR) wavelength range. Experimental results demonstrate the ability of $Ti_3C_2T_x$ MXene to generate stable MIR nanosecond pulses when incorporated into a $Ti_3C_2T_x$ Q-switched erbium-doped ZBLAN fiber laser (EDZFL) operating at 2.8 μm. This indicates the potential of MXene in advancing the design of near-infrared (NIR) and MIR photonic devices [42]. Over the past decade, extensive research has focused on silicon photonic integrated circuits for telecommunication and data centers, with many initiatives approaching commercialization. Recent interest has grown in scaling up the dimensions of silicon-on-insulator (SOI)-based devices to extend the operating wavelength to the short MIR range (2-4 μm). This wavelength range holds great potential for applications in lab-on-chip sensors, free-space communications, and other fields [43].

Another study highlights the infrared radiation characteristics of three MXenes: $Ti_3C_2T_x$, $Ti_3CNT_x$, and $V_4C_3T_x$. The infrared emissivity of a 200 nm thick $Ti_3C_2T_x$ coating is less than 0.06 within the wavelength range of 3-27 μm, while $Ti_3CNT_x$, and $V_4C_3T_x$ exhibit values of 0.13 and 0.26, respectively [44], [45]. Additionally, the low mid-IR emissivity (as low as 10%) of 2D $Ti_3C_2T_x$ MXenes is reported, making them excellent candidates for intrinsic black solar absorbing materials with high spectral selectivity and solar absorptance up to 90%. First-principles calculations suggest that the IR emissivity of MXenes is highly tunable, depending on the orientations of their nanoflakes and terminal groups. Other MXenes, such as $Ti_2CT_x$, $Nb_2CT_x$, and $V_2CT_x$, may also have potential as low-emissivity materials based on these calculations [14]. In [15], an extensive methodology for producing various Zn-based MAX phases and Cl-terminated MXenes is presented. This method involved the replacement reaction of late transition-metal halides with the A-site element in conventional MAX phases. By utilizing molten $ZnCl_2$ and MAX phase precursors ($Ti_3AlC_2$, $Ti_2AlC$, $Ti_2AlN$, and $V_2AlC$), unique MAX phases including $Ti_3ZnC_2$, $Ti_2ZnC$, $Ti_2ZnN$, and $V_2ZnC$ were successfully synthesized. Excess molten $ZnCl_2$, with its strong Lewis acidity, resulted in the exfoliation of $Ti_3ZnC_2$ and $Ti_2ZnC$, producing Cl-terminated MXenes such as $Ti_3C_2Cl_2$ and $Ti_2CCl_2$. This novel approach allows the exploration of non-thermodynamically stable MAX phases at high temperatures, which would be challenging to synthesize using conventional methods. Importantly, it is the first time that exclusively Cl-terminated MXenes have been produced, offering a safe and effective alternative to producing MXenes without using hydrofluoric acid (HF).

In [18], two configurations of 2D $Ti_3C_2T_x$ were utilized to demonstrate two innovative methods for achieving all-optical neuronal nonlinear activation functions. This was accomplished through specific light-matter interactions using a silicon waveguide covered in MXene flakes and a saturable absorber composed of MXene thin film. Furthermore, the development of high-performance mid-IR to deep-UV van der Waals photodetectors using MXenes was reported in [21]. Additionally, a 2D titanium carbide ($Ti_3C_2T_x$) MXene wideband plasmonic metamaterial absorber was created [22]. This absorber exhibited potent localized surface plasmon resonances in the near-infrared frequencies, enabling high-efficiency absorption (90%) over a large wavelength window. The synthesis, characterization, and applications of MXenes, particularly 2D carbides, and nitrides, have gained significant research attention over the past decade. These materials have been explored for energy storage, adsorbents, solid-state laser absorbers, membrane materials, photothermal tumor control bio-windows, antibacterial agents, and biomedical applications. MXenes such as $Nb_2CT_x$ and $V_2CT_x$ have shown potential as energy storage materials, while $V_2CT_x$ has demonstrated electrocatalytic properties for $N_2$ reduction. Additionally, MXenes like $Ti_3C_2T_x$ have garnered interest in the development of advanced thermal camouflage technologies due to their exceptionally low mid-IR emissivity. The structural diversity of MXenes has expanded over the years, with the synthesis of $M_2X$, $M_3X_2$, $M_4X_3$, and $M_5C_4$.

Due to the distinct electrical structure and enormous specific area, 2D materials are used in energy storage and conversion. An efficient and readily available electrocatalyst for the electrocatalytic $N_2$ reduction reaction (NRR) is made and examined in this study as 2D vanadium carbide ($V_2CT_x$ MXene). With a Faradaic efficiency of 4% at -0.7 V as compared to a reversible hydrogen electrode, this electrocatalyst produces a significant amount of $NH_3$ when tested in 0.1 M $Na_2SO_4$. Theoretical calculations indicate that this catalyst has a low reaction barrier of 0.88 eV along the distal channel [46]. Here in [42] focused on MXene Photonic Devices for Near-Infrared to Mid-Infrared Ultrashort Pulse Generation. In this study [47], 2D $Ti_3C_2T_z$ MXene Synthesized by Water-free Etching of $Ti_3AlC_2$ in Polar Organic Solvents. In this study [48] investigated the applications of Few-Layer $Nb_2C$ MXene: Narrow-Band Photodetectors and Femtosecond Mode-Locked Fiber Lasers. Under equilibrium conditions, stoichiometric vanadium carbide (VC) cannot be produced. It often crystallizes as a disordered NaCl cubic -$VC_{1-x}$ ($VC_{0.65}$-$VC_{0.90}$) with significant homology. Under equilibrium conditions, stoichiometric vanadium carbide (VC) cannot be produced. It often crystallizes as a disordered NaCl cubic -$VC_{1-x}$ ($VC_{0.65}$-$VC_{0.90}$) with significant homology. There are structural flaws because the carbon atoms in NaCl-type vanadium carbides can only partially fill the octahedral vacancies of the metal face-centered cubic (FCC) sublattice. Their presence may occasionally result in atomic ordering, which is brought on by structural voids at interstitial lattice sites and the redistribution of nonmetallic atoms. This non-stoichiometric interstitial compound can be exploited in the field of electronic materials because of its high concentration of structural vacancies. Like VC, $V_4C_3$ has 7 atoms total of 4 V atoms and 3 C atoms, and one C atom vacancy in each cell [49]. Niobium carbide ($Nb_2C$), a newly discovered MXene family member, has recently attracted interest in several disciplines, including materials science, physics, chemistry, and nanotechnology. The material characteristics, synthesis techniques, and applications of $Nb_2C$ in energy storage, catalysis, and photovoltaic devices [50].

**Two-Dimensional MXenes and Types of MAX Structures**

MXenes are a class of two-dimensional (2D) materials consisting of transition metal carbides, nitrides, or carbonitrides found in layered structures [35], [51], [52]. These materials have gained significant attention as functional building blocks for optoelectronic and photonic devices following the discovery of stable 2D atomic carbon layer graphene. MXenes offer distinct advantages over their bulk counterparts due to their unique structures. The general formula for MXenes is $M_{n+1}X_nT_x$, where M represents an early transition metal (such as Mn, Mo, Cr, Ta, Nb, V, Hf, Ti, Sc, Zr), X denotes carbon or nitrogen, T represents surface functional groups (such as fluorine, oxygen, or hydroxyl), and n can be 1, 2, or 3 [53]. The MXene $Ti_3C_2T_x$, for example, is typically prepared through acid etching methods [54]. Since the discovery of $Ti_3C_2T_x$ in 2011, MXenes has attracted considerable attention as a diverse family of 2D transition metal carbides and nitrides. The unique combination of metallic conductivity and hydrophilicity in $Ti_3C_2T_x$ has led to remarkable advances in electrochemical applications. Surface terminations, such as hydroxyl, oxygen, or fluorine, continuously form on the chemically active surface of MXenes, influenced by the choice of etching chemicals. These surface functional groups impact the hydrophilicity, electrochemical characteristics, electronic structure, conductivity, work function, and overall electronic properties of MXenes [55].

Niobium-based MXenes, such as $Nb_2CT_x$ and $Nb_4C_3T_x$, possess distinctive properties and hold promise for various applications. MAX phases, which are layered ternary carbides/nitrides with a hexagonal structure, serve as precursor phases for MXenes. MXenes are produced by selectively etching the A-site atoms in these MAX phases to create their 2D counterparts. MXene materials exhibit a hydrophilic surface and metallic conductivity, making them highly attractive for applications in photocatalysis and energy storage. The structural arrangement of MAX phases is illustrated in **Fig. 3** [51], [53], [56], [57].

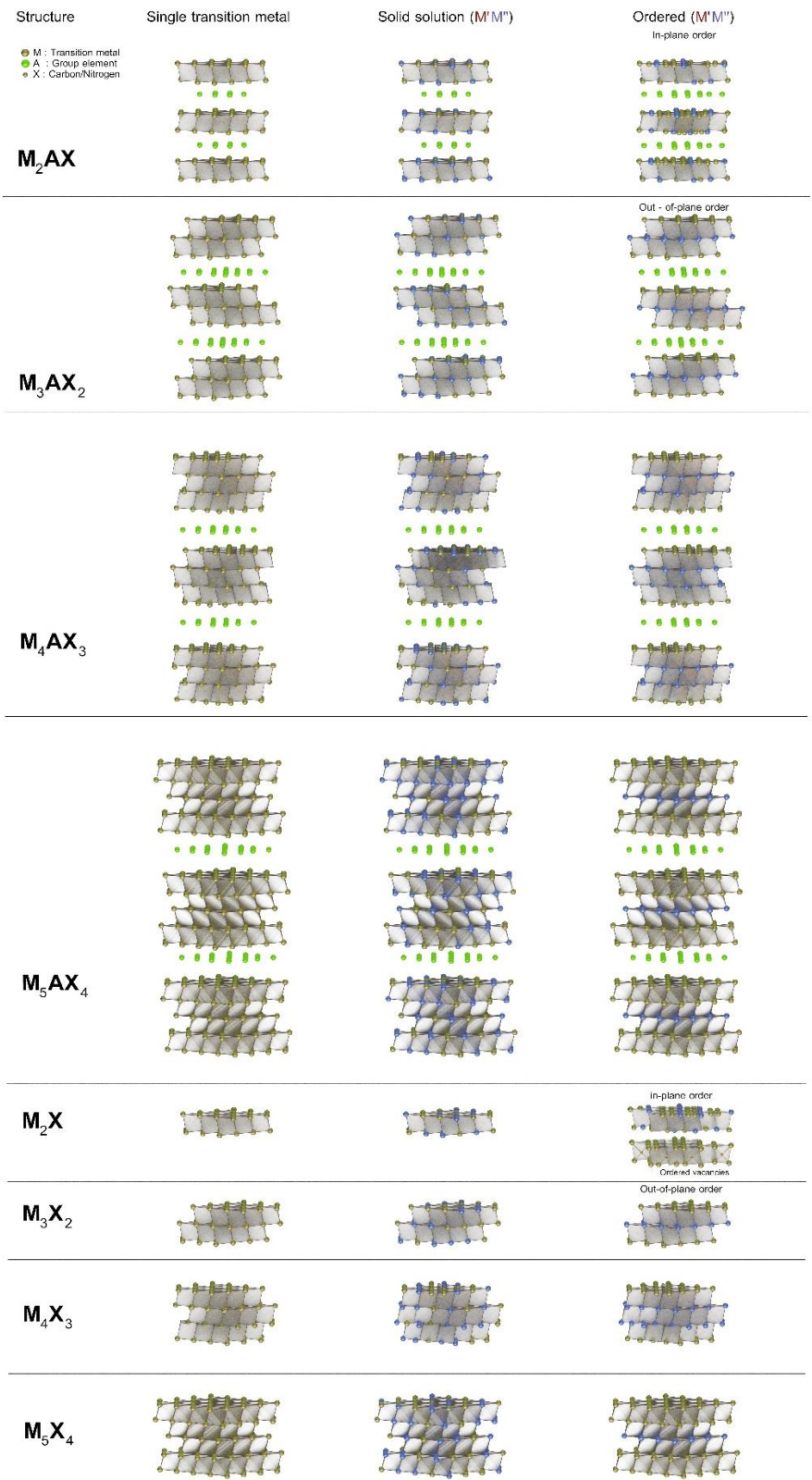

**Figure 3.** The MAX's phase structure is depicted schematically, along with the matching MXene. The figure focuses on the in-plane and out-of-plane order of the crystal orientation of the atoms [56].

**Mxene Material Synthesis and Process**

The synthesis of 2D materials has been approached using two main methods: bottom-up and top-down   [35], [58]. Chemical vapor deposition (CVD), commonly employed to produce graphene, is an example of the bottom-up approach. However, CVD has limited applicability in MXene synthesis as it typically results in multilayer films rather than single-layer MXenes. For instance, CVD has been used to produce $Mo_2C$ films consisting of at least six layers, rather than single-layer MXenes. While CVD has been successful in obtaining other 2D structures like $MoS_2$, further research is needed to explore its potential in MXene production. In recent years, significant interest has been directed towards 2D materials due to their exceptional mechanical, thermal, electrical, and optical properties. MXenes, as a class of innovative 2D materials, has garnered substantial attention with more than 30 different types of MXenes being successfully synthesized. Ongoing research aims to discover additional MXenes and explore their potential as high-performance optoelectronic materials. This article primarily focuses on the production of MXenes, their optical characteristics, and their applications in ultrafast photonics **(Fig. 4)**. Two main methods used to produce MXenes are aqueous acid etching (AAE) and chemical vapor deposition (CVD) [59].

The top-down strategy for MXene synthesis can be divided into mechanical and chemical exfoliation, as shown in **Fig. 3**. Selective etching is employed to dissolve the strong covalent bonds between the A and MX layers in the MAX phase. Etching is primarily achieved using molten salts and hydrofluoric acid (HF). In this process, the M-A metal bond is replaced by hydroxyl, oxygen, or fluorine. The HF method for producing 2D MXenes involves two main steps: exfoliation and etching. However, direct HF application is environmentally hazardous and poses health risks. To address this, an alternative approach involves using a less harmful weak acid that interacts with a fluorinated salt to generate in-situ HF. This technique, known as the molten salt method, utilizes fluorinated Lewis salts, eliminating the use of fluoride in the synthesis process. The synthesis steps in molten salts closely resemble those in conventional HF methods. High-temperature molten salts such as $ZnCl_2$ and $CuCl_2$ can effectively strip a wider range of MAX-phase materials, with Lewis's acids in the molten salt acting as F- and H+ equivalents. Electrochemical etching, minimum intensive layer delamination (MILD), and MAX etching are also employed to produce high-quality, non-toxic MXenes [35].

The initial mechanical exfoliation technique, using sticky tapes, was employed to isolate graphene. However, the strong covalent or metallic bonds between the M and A (typically Al) layers in MAX phases make selective removal by mechanical forces challenging. Consequently, chemical exfoliation is predominantly used to obtain MXenes from their MAX phase precursors. The core idea behind this approach is to etch away the relatively strong interlayer bonds, a topic that will be further explored in the following section along with precursor selection and subsequent exfoliation into single flakes [58]. Liquid-phase exfoliation is the commonly employed method for MXene production, as depicted in **Fig. 4**. This method involves progressive delamination and selective etching of the metallic "A" layer in MAX powder, which typically contains Al, Ga, or Si.

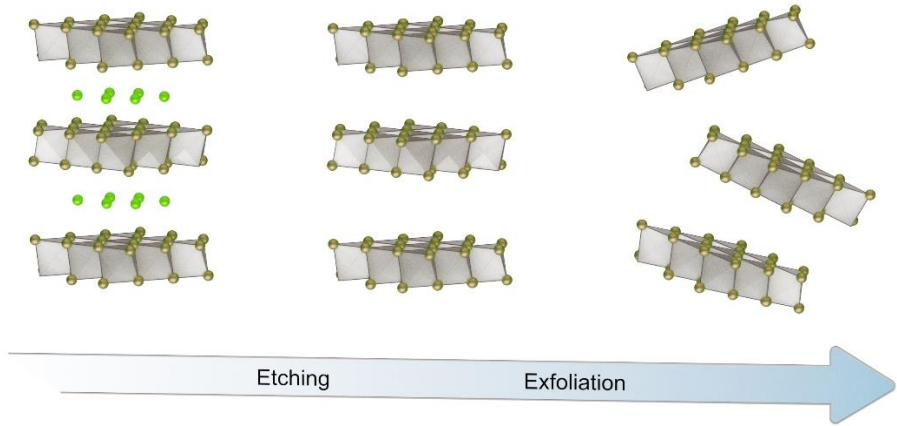

**Figure 4.** Synthesis pathway of 2D MXene flakes. First, selective etching is performed on MXenes, and the product is then exfoliated to form single flakes.

Several bottom-up synthesis techniques have been developed for MXenes, including chemical vapor deposition, template-based synthesis, and plasma-enhanced pulsed laser deposition. These methods enable the production of MXenes with controlled structures, sizes, and good crystalline quality. However, the stability of MXenes has been a limiting factor, which researchers have been working to improve. Mild reaction conditions are preferred as high concentrations of hydrofluoric acid (HF) can accelerate the degradation and alter the structure of MXenes. The use of organic solvents helps reduce oxidation while minimizing contact with water is crucial to prevent oxidation. MXenes tend to oxidize more rapidly in liquid environments compared to solid ones, and processes such as photocatalysis and thermo-catalysis can accelerate this degradation process.

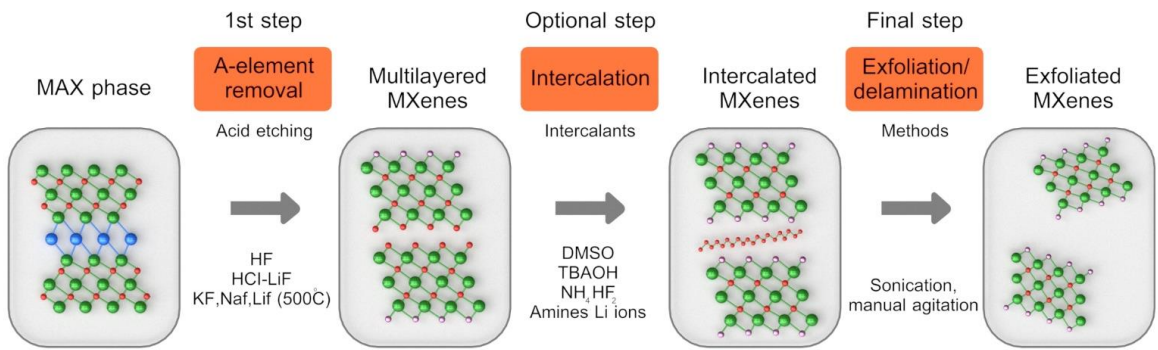

**Figure 5.** Liquid-phase processes for MXene synthesis and characterization of synthesized MXenes.

In bottom-up synthesis, small organic or inorganic molecules or atoms serve as the fundamental building blocks. These units can then assemble to form 2D-ordered layers through crystal formation and organization. Chemical vapor deposition is a prominent technique for this approach as it allows the production of high-quality thin films on various substrates. For example, high-quality ultrathin $Mo_2C$ films were synthesized using chemical vapor deposition with methane gas as the carbon source and a Cu/Mo foil as the substrate at temperatures above 1085°C. By optimizing the growth temperature and time, films with lateral diameters of approximately 50 nm were achieved. The faultless nature and high degree of crystallinity of these $Mo_2C$ films indicate the absence of surface functional groups. However, for biomedical applications that require well-functionalized MXenes for surface engineering of nanosheets, the large lateral diameter of 50 nm may hinder cell penetration. In addition to chemical vapor deposition, the template method and plasma-enhanced pulsed laser deposition have also been investigated for MXene synthesis **(Fig. 5)** [35].

Highly selective etching can be achieved using hydrofluoric acid (HF), HF-containing etchants (such as $NH_4HF_2$ salt), or HF-forming etchants (such as LiF salt + HCl). Various liquid-phase techniques have been developed to synthesize more than 30 MXenes, allowing for the preparation of well-suited MXene films or flakes for different applications. For instance, for transparent conductive electrodes (TCE), highly conductive MXene films are required, while electronic devices with MXene as an intermediary layer at the metal-semiconductor junction necessitate work-function-modulated MXene flakes with controlled functional groups. Porous MXene films find application in energy storage systems such as lithium-ion, sodium-ion, or lithium-sulfur batteries. Achieving the desired conductivity, electrical characteristics, thickness, size, or morphology of MXene relies on selecting the appropriate etchant, delamination technique, and intercalants. In high-performance photodetectors or photovoltaic devices, highly conductive and work-function-regulated MXenes are essential for use in contact electrodes. Patterned MXene films play a crucial role in creating the plasmonic effect for optoelectronic applications. On the other hand, semiconducting MXenes, with their tunable bandgap through functional group modifications, are suitable for the sensing channel in broadband detection photodetectors. This summary provides an overview of the general liquid-phase MXene synthesis process and recent studies focused on developing suitable MXene films for nanophotonic applications.

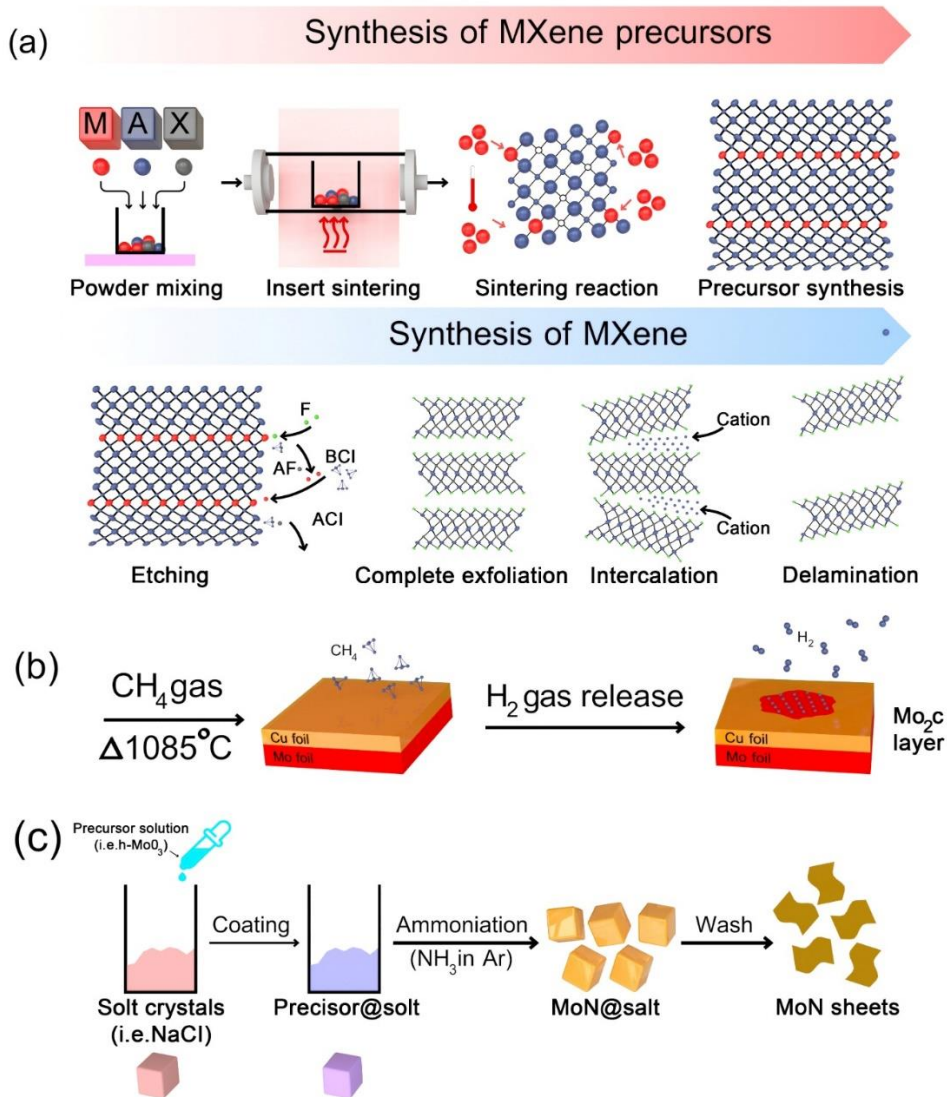

**Figure 6.** (a) Top-down steps in the synthesis of MXene nanomaterials from precursor to etching; (b) chemical vapor deposition of Mo and C to produce a thin film of Mo₂C in a gas chamber; and (c) the synthesis of MoN nanosheets using a salt template. (b) Chemical vapor deposition of Mo and C to generate a thin film of Mo₂C. Reproduced from references with permission.

     In recent years, research efforts have explored MXene synthesis methods without the use of solvents or etchants, resulting in large-grain size and highly crystalline MXene films that are free from intercalation molecules and exhibit -F, -Cl, or -OH groups. These investigations aim to develop high-quality MXene flakes for electrical and optical applications that are free from contamination and defects. Inkjet printing has also garnered significant interest as a scalable approach with potential for real-world nanophotonic applications [60]. The production of MXene begins with precise control of temperature, hydrofluoric acid (HF) concentration, and etching time during the treatment of its MAX precursor, such as Ti₃AlC₂, in an aqueous HF solution. This method has become the primary approach for MXene synthesis due to its simplicity, cost-effectiveness, and high yield **(Fig. 6)**. The key to separating the atomic layers of "A" from

the ternary MAX phase lies in the weaker metallic bond between M and A compared to the mixed covalent/metallic/ionic bond between M and X. HF acid has been successfully employed to produce over 20 different MXenes. During the exfoliation process, A-elements react with HF, forming fluorides (such as $AlF_3$, and $SiF_4$), and gaseous hydrogen ($H_2$), resulting in the accordion-like structure of $M_{n+1}X_n$. So far, only Al and Si have been etched to create MXenes, despite the identification of more than 10 other types of A-elements in MAX phase precursors. The etching conditions for $M_2AX$ phases are typically milder than for $M_3AX_2$ and $M_4AX_3$ phases, and MAXs containing Al generally have fewer valence electrons compared to their Si counterparts. Wet acid etching has proven to be a simpler method for manufacturing carbides compared to nitrides or carbon nitrides. Wet acid etching has been successful in generating only a few nitrides to date, and none of them belong to the $M_4X_3$ phase. Aggressive or prolonged etching can lead to the destruction of MXenes and the formation of carbide-derived carbon.

After selective etching, the dangling metal linkages on the MXene surface become unstable and readily react with aqueous solutions **(Eq. (1) and (2))**, resulting in surface terminations such as -O, -OH, and -F. These unavoidable surface terminations contribute to the intrinsic hydrophilic nature of MXenes and significantly influence their electrical, optical, magnetic, and mechanical properties [61].

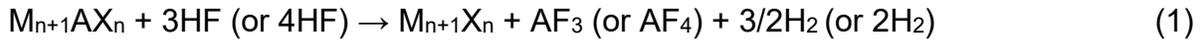
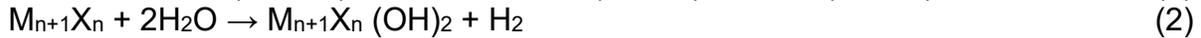
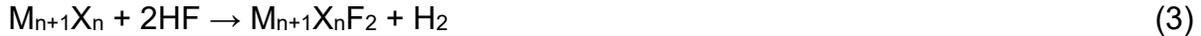

$$M_{n+1}AX_n + 3HF \text{ (or } 4HF) \rightarrow M_{n+1}X_n + AF_3 \text{ (or } AF_4) + 3/2H_2 \text{ (or } 2H_2) \qquad (1)$$
$$M_{n+1}X_n + 2H_2O \rightarrow M_{n+1}X_n (OH)_2 + H_2 \qquad (2)$$
$$M_{n+1}X_n + 2HF \rightarrow M_{n+1}X_nF_2 + H_2 \qquad (3)$$

## Classification of MXenes Structure

### I. Mono Transition Metal MXenes

MXenes are derived from the parent MAX phases and exist in three forms: $M_4C_3$, $M_3C_2$, and $M_2C$. These MXene forms inherit their structures from the MAX phases, which consist of an element or layered precursor such as $Mo_2Ga_2C$ with the general formula $M_{n+1}AX_n$, where X can be either C or N (n = 1-4), A belongs to groups 13 or 14, and M represents an early transition metal. The MAX phases exhibit a layered hexagonal structure with nearly closed-packed M layers and X atoms occupying octahedral positions. Interspersed with the $M_{n+1}X_n$ layers is the A element, which is metallically bonded to the M element **(Fig. 7a)** [35].

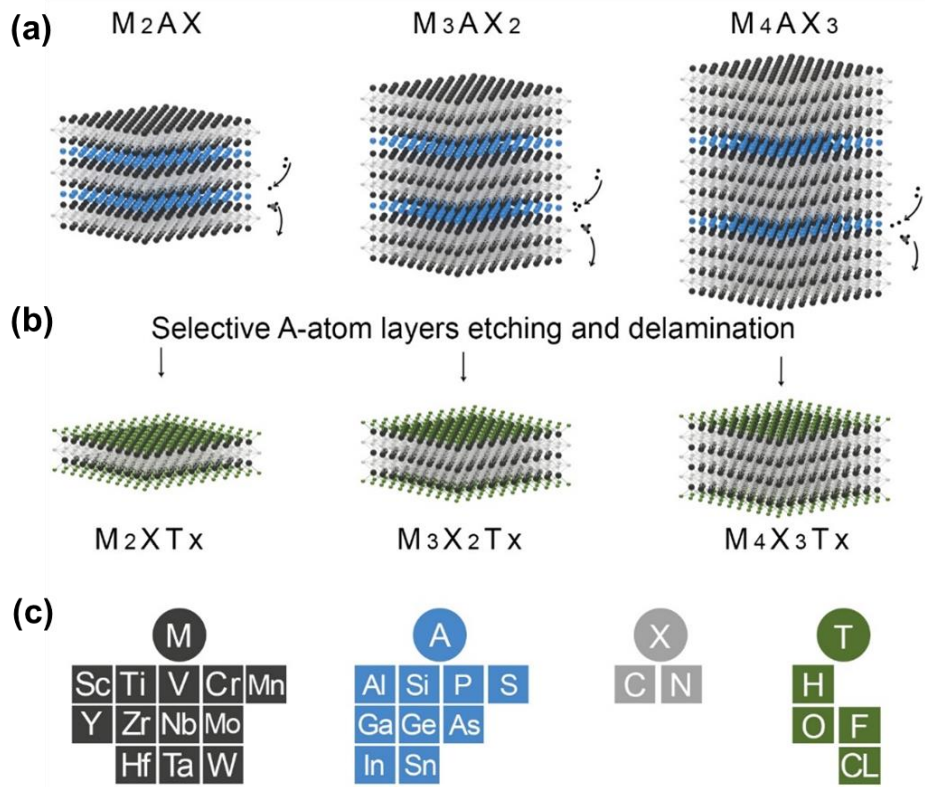

**Figure 7.** (a) After the production of surface terminations (yellow atoms) and the A-group layers were etched selectively (red atoms), three different mono-M MAX phases—$M_2AX$, $M_3AX_2$, and $M_4AX_3$—MXenes were produced. M, A, X, and T components that could be present in the MAX and MXene phases; (b) the synthesized $M_2X$, $M_3X_2$, and $M_4X_3$ MXenes, which are double transition metals (DTM). The transition metals M′ and M″ are represented by the green and purple elements, correspondingly. Both the in-plane divacancy order (M′4/3X) and the in-plane order (M′4/3M″2/3X). The order is out of a plane (M′2M″X2 and M′2M″2X3). M′ and M″ transition metals are dispersed throughout the disordered MXenes in solid solutions. Additionally, solid solution $(Mo_{0.8}V_{0.2})_5C_4T_x$ was prepared successfully; however, it has not been shown for simplicity's sake. Reproduced with permission from reference.

## II. Double Transition Metal MXenes

Double transition-metal (DTM) MXenes differ from their mono-transition-metal counterparts in that they feature metal sites occupied by two transition metals. In ordered DTM MXenes, the transition metals are arranged in either an in-plane or an out-of-plane ordered framework. The defining characteristic of DTM MXenes is the random distribution of two transition metals throughout their 2D structure **(Fig. 7b)**. This unique structural diversity and the variety of transition-metal pairings in DTM MXenes enable customization for specific properties such as magnetic, optical, thermoelectric, electrochemical, mechanical, and catalytic applications. The ability to precisely control the structure and composition of DTM MXenes represents a significant advancement in the realm of true 2D materials, opening new avenues for the development of tailored nanoparticles for specific purposes [35].

The synthesis of nearly 30 MXenes has primarily involved etching MAX phases containing Al layers. MAX phases are characterized by layered and hexagonal structures that exhibit both metallic and ceramic properties. These phases are composed of early transition metal carbides and nitrides. They are categorized into three main groups based on the number of M layers between the A layers: "211" ($M_2AX$), "312" ($M_3AX_2$), and "413" ($M_4AX_3$). The current MAX group encompasses over 130 distinct structures, the majority of which crystallize in the P63/mmc space group and its derivatives.

**Etching techniques**

Etching is a crucial step in the synthesis of MXenes, as it involves the substitution of space termination groups (-OH, -O, -F) for the A-group element layer. This process leads to the formation of an $M_{n+1}X_nT_z$ multilayer structure held together by weak hydrogen and/or van der Waals bonds. The most common method for removing the A layers from MAX phases utilizes acidic aqueous solutions containing fluoride. The specific etching parameters, such as HF concentration and duration, are influenced by the chemistry and structure of the parent material. For example, $Ti_3C_2T_z$ can be obtained from $Ti_3AlC_2$ using 50 wt.% HF, while $Ti_2CT_z$ is produced with 10 wt.% HF and $Ti_2AlC$ are fully dissolved with 50 wt.% HF. The synthesis of MXenes with larger n values generally requires longer etching times.

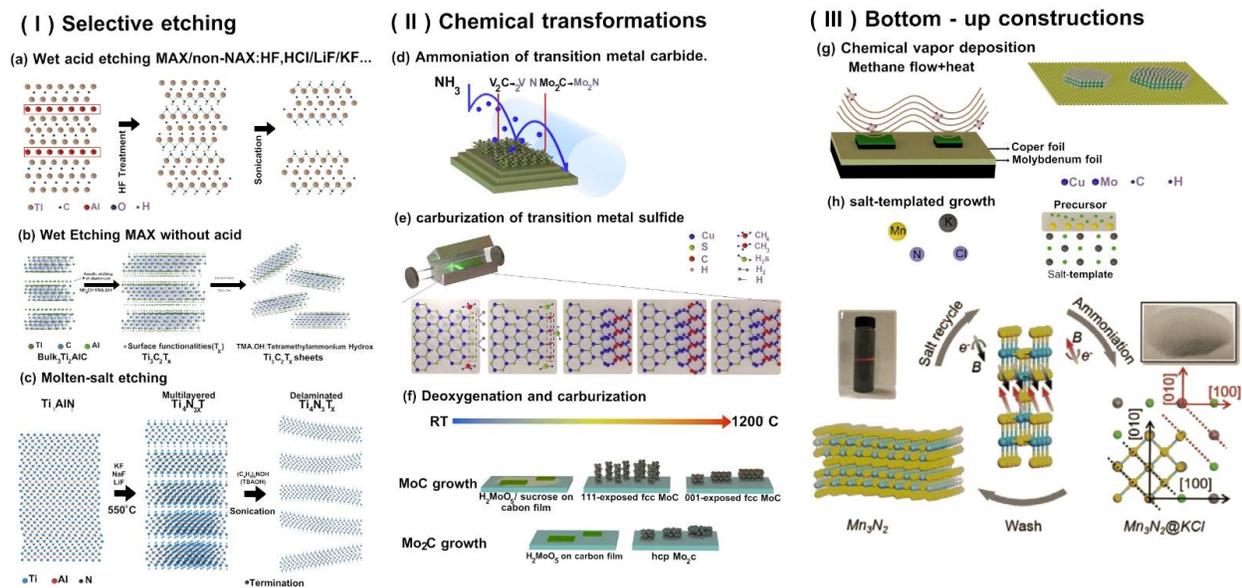

**Figure 8.** Schematic of the MXene synthesis techniques [61].

To mitigate the hazards associated with HF, an alternative approach involves generating HF on-site by combining a strong acid with a fluoride salt. Previous studies have demonstrated successful etching of $Ti_3AlC_2$ using a solution of lithium fluoride (LiF) and hydrochloric acid (HCl). It has been found that the presence of metal halides is necessary for the creation of MXene "clay." Other etchants such as HCl, NaF, KF, and $NH_4F$ have also been used with varying success. Hydrothermal synthesis in the presence of molten salts such as LiF, NaF, and KF has been employed for the synthesis of $Ti_3C_2T_z$ and the first nitride MXene, $Ti_4N_3T_z$. The Lewis acidic molten salt $ZnCl_2$ has recently been used to synthesize completely Cl-terminated MXenes ($Ti_3C_2Cl_2$ and $Ti_2CCl_2$). Carbide MXenes can be converted to nitride MXenes through ammonion. For instance, $V_2CT_z$ and $Mo_2CT_z$ can be transformed into $V_2NT_z$ and $Mo_2NT_z$, respectively, by ammonion at 600°C. Wet chemical etching has successfully produced nitride MXene $Ti_2NT_z$. The etching techniques involving acidic fluoride-containing solutions are typically carried out at temperatures below 60°C to avoid the formation of carbon from carbides. However, higher temperatures can be employed to synthesize MXenes. For example, $Ti_4N_3T_z$ was synthesized at 550°C in molten salts, and $Ti_3C_2T_z$ was created using a hydrothermal process with NaOH at 270°C [58].

**Exfoliation Process**

To obtain colloidal suspensions consisting of one or a few MXene layers, the multilayers produced by etching need to undergo exfoliation or delamination. Before exfoliation, it is important to rinse the multilayers in water or acidic solutions such as HCl or sulfuric acid ($H_2SO_4$) to remove any residual salts from the etching process. The choice of exfoliation technique depends on the specific etching conditions.

When using HCl/HF etchants, the presence of lithium cations and water intercalates between the MXene layers, creating more space and weakening their interactions. This facilitates natural exfoliation when the pH of the rinsed multilayers reaches approximately 6. Simply shaking or sonicating the material can increase the yield of exfoliation. However, direct sonication alone typically leads to poor yields. Therefore, for other etchants, exfoliation is often achieved through the intercalation of cations or organic molecules that reduce interlayer interactions and increase the spacing between layers. Subsequently, shaking or sonication can be employed.

The selection of the exfoliation process also depends on the MXene composition. For example, only $Ti_3C_2T_z$ and $(Mo_{2/3}Ti_{1/3})_3C_2T_z$ can be effectively exfoliated using dimethyl sulfoxide (DMSO) as the molecule. Tetrabutylammonium hydroxide (TBAOH) can be used to treat several MXenes, including $V_2CT_z$, $Ti_3CNT_z$, $(Mo_{2/3}Ti_{1/3})_3C_2T_z$, $(Mo_{1/2}Ti_{1/2})_4C_3T_z$, $Ti_4N_3T_z$, and $Mo_2CT_z$, while $Ti_3C_2T_z$ requires the use of tetramethylammonium hydroxide (TMAOH) as an alternative. After exfoliation, centrifugation can be employed to obtain a colloidal suspension consisting of only one or a few layers of MXene. If the product is not immediately used, it should be stored close to pH neutral. Other processing techniques for these ultrathin sheets include vacuum filtration, spin coating, acid or base crumpling, and electrophoretic deposition [58].

**Coating Techniques of MXenes**

In the past decade, there has been significant interest in coating MXene dispersions and utilizing printing techniques to develop MXene-based designs and devices. Direct coating techniques have emerged as promising and versatile methods for producing thin and continuous conductive sheets or electrodes. These techniques include vacuum-assisted filtering, stamping, spray coating, dip coating, and spin casting, among others. These approaches enable the creation of high-performance functional electronics with advanced features such as stretchability, flexibility, and foldability, using the thin and conductive coatings obtained from MXene.

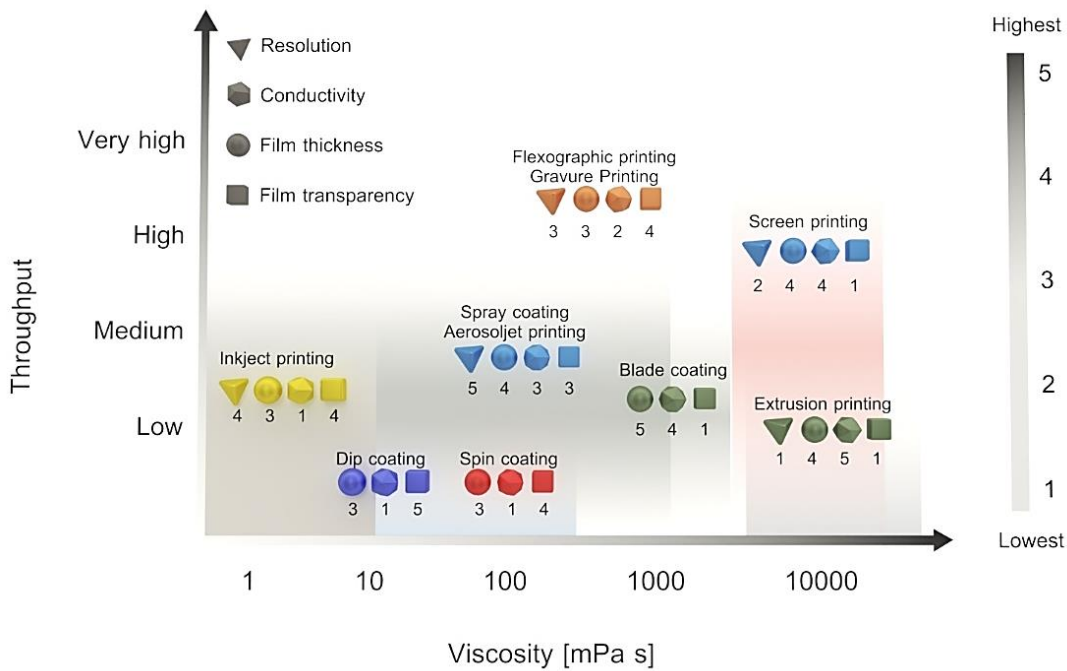

**Figure 9.** The flow chart for processing colloidal MXenes in solutions. For the most common printing/coating techniques, a relationship between the throughput, ink viscosity, resolution (only pertains to printing methods), film conductivity, film thickness, and film transparency is shown in [62].

## I. Dip-coating

Dip coating offers two advantages: ease of use and the potential for scalable manufacturing of large-area thin films. When the viscosity of MXene dispersions is low, dip coating is recommended for producing a range of MXene films, from highly transparent films with good conductivity to moderately thick films. MXene colloidal dispersions typically exhibit negative zeta potentials (ranging from 30 to 80 mV), which promote repulsive interactions between the negatively charged MXene nanosheets. This stability leads to homogeneous dispersions suitable for dip coating. The dip coating process can be generally divided into four stages: dipping, immersion, deposition and drainage, and solvent evaporation, as illustrated in the recommendation map **(Fig.9)** [62].

## II. Spray-coating

Spray coating is a direct method that allows to production of large-area MXene films, particularly when the viscosity of the MXene dispersion is relatively low. Traditionally, delaminated MXene aqueous dispersions have been used for spray coating to create conductive thin films. However, there is a lack of reports on non-aqueous MXene dispersions, except for ethanol-based colloidal solutions. This is because only polar organic media with Hildebrand and Hansen parameters that are compatible with MXenes are capable of effectively dispersing MXene flakes in non-aqueous solvents. While these polar organic solvents can disperse the flakes well, they often have high boiling points, making it challenging to achieve solvent evaporation at low temperatures. This limitation restricts the film thickness and process efficiency.

## III. Spin-coating

Spin coating is a highly recommended method for processing MXene dispersions when their viscosity falls within the medium range (10-200 mPas) or their concentration is in the medium range (120 mg/mL). Among solution-processing techniques, spin coating is the most widely used approach for creating thin and uniform films from dispersions. In this process, the dispersion is initially spread over the substrate as it wets the surface. As the substrate is spun, the excess liquid is flung off, leaving behind a thin liquid layer that rapidly undergoes solvent evaporation. This results in the formation of homogeneous thin films. The thickness of the film depends on various factors such as MXene concentration, spin speed, and duration. For example, at a given spin speed, using a higher concentration of MXene dispersion (>10 mg/mL) produces thicker films compared to more diluted dispersions (1 mg/mL). The spin coating enables the fabrication of films with thicknesses ranging from nanometers to micrometers and varying transmittance, making it particularly suitable for lab-scale prototype devices with flat substrate surfaces.

## IV. Blade-coating

Blade coating is another solution-processing technique suitable for the mass production of MXene films, particularly when working with higher concentrations (20-200 mg/mL) or viscosities (up to 2000 mPa·s) of MXene dispersions. Like spin coating, blade coating can produce large-area MXene films with well-aligned flakes from concentrated dispersions. These films exhibit excellent electrical conductivity and mechanical properties, including high tensile strength and Young's modulus. The thickness of the MXene films produced through blade coating typically ranges in millimeters, depending on the blade height and MXene concentration. Concentrated MXene dispersions that have undergone delamination (i.e., exhibiting an apparent viscosity greater than 1 Pa·s) are known to exhibit shear-thinning behavior, allowing for good flowability and processability during blade coating.

## Properties of Mxenes

## I. Surface Functional groups

MXene surfaces exhibit hydrophilicity with water contact angles ranging from 21.5° to 35°, which is attributed to the presence of surface terminal groups. Moreover, MXenes generally possess high conductivities, with $Ti_3C_2T_x$ being the most extensively studied MXene and exhibiting a conductivity of $10^4$ $Scm^{-1}$. This unique combination of hydrophilicity and metallic conductivity sets MXenes apart from graphene and its derivatives. The substantial surface area inherited from MXenes' 2D structure makes them promising candidates for electrochemical energy storage applications. The electronegative O or F atoms in the terminal groups give rise to negatively charged surfaces in functionalized MXenes. This allows for reversible electrochemical intercalation of cations such as $Na^+$, as well as multi-valence cations like $Mg^{2+}$ and $Al^{3+}$, through strong Coulombic interactions with the MXene surface. The surface groups also play a significant role in the pseudocapacitive redox process, where oxygen groups can be converted into hydroxyl groups while accommodating both protons and electrons in $H_2SO_4$ electrolytes.

MXenes exhibit stable dispersion behavior in polar solvents such as water, ethanol, acetone, acetonitrile, N, N-dimethylformamide (DMF), dimethyl sulfoxide (DMSO), N-methyl-2-pyrrolidone (NMP), and propylene carbonate (PC), enabling easy integration with various solvent-sensitive materials. MXene electrodes, such as $Ti_3C_2Tx$, have demonstrated excellent electrochemical performance in supercapacitors with macroporous, hydrogel, and vertically aligned liquid crystal structures. These electrodes exhibit ultrafast operating rates of 100 $Vs^{-1}$ and thickness-independent capacitance.

The surface functional groups of MXenes not only impact their electrochemical behavior but also influence their electrical and optical characteristics, including band structure and work function. The specific species of surface functional groups are dependent on the synthesis method and post-synthesis treatments. Quantification of $Ti_3C_2Tx$ MXenes produced via the HF approach and the HCl-LiF route has been successfully carried out using 1H and 19F nuclear magnetic resonance (NMR) spectroscopy [55].

## II. Tunable Interlayer Spacing

MXenes offer the ability to adjust the interlayer spacing through the manipulation of intercalated species between the nanosheets. In the case of $Ti_3C_2T_x$ MXene, it exhibits hygroscopic properties and reversible humidity-dependent behavior once cation species are intercalated, resulting in interlayer spacing of 12.5-15.5 (corresponding to one or two intercalated water layers). Furthermore, organic molecules can also intercalate between MXene nanosheets. By using cetyltrimethylammonium bromide (CTAB) or stearyl trimethylammonium bromide (STAB), $Ti_3C_2T_x$ with a pillared structure and interlayer spacings of 22.3 or 27.1, respectively, have been observed. Various alkylammonium cations have been intercalated, allowing for tunable interlayer spacing ranging from 14.7 to 38.0. For example, the interlayer spacing of tetrabutylammonium cation ($TBA^+$) intercalated in $Mo_2CT_x$ varies from 16.9 to 18.9 depending on the moisture level. These variable interlayer spacings hold potential for applications such as gas separation and sensing [55].

## III. Band Structure

Extensive theoretical investigations have been conducted to explore the electronic band structures and intriguing properties of MXenes within their diverse family. While a few MXene systems exhibit semiconducting behavior, the majority of functionalized MXenes are predicted to possess metallic or semi-metallic band structures. Band gaps ranging from 0.24 to 1.8 eV have been observed in $Sc_2CT_2$ (T = O, OH, F), $Ti_2CO_2$, $Zr_2CO_2$, and $Hf_2CO_2$. Notably, biaxial strains of 4%, 10%, and 14% can induce a transition from an indirect to a direct band gap in $Ti_2CO_2$, $Zr_2CO_2$, and $Hf_2CO_2$, respectively. The formation of newly generated states below the Fermi level is attributed to strong band hybridization between the transition metal atoms and the carbon or oxygen atoms. This hybridization leads to the opening of the band gap as MXenes undergo functionalization, facilitated by the lower electronegativity of the transition metals compared to the functional groups and carbon atoms. Ti, Zr, and Hf MXenes with O group functionalization exhibit similar metallic to semiconducting transition trends due to their shared outer-shell electronic configuration. As the atomic number of the metal increases, resulting in reduced electronegativity, the band gap of $M_2CO_2$ (M = Ti, Zr, Hf) MXenes also increases.

Among MXenes, only a small subset containing two different transition metal atoms exhibits an ordered structure instead of a random solid solution. These ordered double transition metal MXenes, with general formulas $M'_2M''C_2$ and $M'_2M''_2C_3$, have M' forming the outer surfaces and M'' forming the central layers. The electronic properties of MXenes are influenced by this ordered structure, as exemplified by the contrasting electronic behavior of $Ti_3C_2T_x$ (metallic) and $Mo_2TiC_2T_x$ (semiconducting). Oxygen-terminated double transition metal MXenes, such as $M'_2M''C_2O_2$ and $M'_2M''_2C_3O_2$ (M' = Mo, W; M'' = Ti, Zr, Hf), have been predicted to exhibit topological semimetal properties. The $Mo_2M''C_2O_2$ (M'' = Ti, Zr, Hf) system demonstrates a significant topological gap that could enable the quantum spin Hall effect at ambient temperature. MXenes with oxygen-terminated surfaces are resistant to oxidation in ambient air. Additionally, certain MXenes, including $M_2CO_2$ (M = W, Mo, Cr), have been identified as topological insulators. Experimental synthesis of these topologically non-trivial MXenes has been achieved in some cases, positioning them as potential candidates for research in topological superconductivity, as well as applications in electrical and spintronic devices [55].

## IV. Electronic properties

Theoretically, numerous studies have been carried out to conduct a thorough investigation of the band structures, density of states (DOS), and many other electrical features. First-principal calculations and experimental findings have thus far shown that the majority of (functionalized) MXenes should be metallic or semi-metallic, while semiconductor MXenes only take a tiny number [55]. Due to the outer layer of transition metal elements, non-terminated MXenes often have a high density of states (DOS) at the Fermi surface. The d-electrons of transition metals dominate the DOS close to the Fermi surface, whereas the p-electrons of X atoms create energy bands 3 to 5 eV below the Fermi surface. It is hypothesized that the electrical characteristics of MXenes' outer transition metal layers are more significant than those of their inner transition metal layers. As a result, surface terminations that are bonded to the transition metal atoms in the outer layer might drastically change the electronic band structures. The Fermi surface is spontaneously lowered or the DOS density at the Fermi surface is reduced when the electronegative termination accepts one or two electrons from the surrounding transition metal layers and generates a new energy band below it. The electrical structure of MXenes is affected similarly by OH- and F-groups since they can only accept one electron, however, an O-group may accept two electrons and so has a distinct effect. According to the results of the initial MXene studies, $Ti_3C_2$ is a metal with a zero bandgap and a finite density of states at the Fermi level. Surface functionalization with OH and F causes the bandgap to be broken, resulting in bandgap values of 0.05 eV and 0.1 eV, respectively **(Fig. 9a)**. Electric fields or external strain can be used to engineer the band gaps of $Ti_2CO_2$ and $Sc_2CO_2$. The electrical and thermal transport of MXenes is similarly impacted by the surface functional groups. The highest electronic transmission is specifically seen in F-terminated MXenes, whereas surface functionalization with O atoms significantly reduces the electronic transmission. In addition, the surface terminations can induce nearly free electron (NFE) states near the Fermi level in various MXenes, e.g., $Ti_2C(OH)_2$, $Zr_2C(OH)_2$, $Zr_2N(OH)_2$, $Hf_2C(OH)_2$, $Hf_2N(OH)_2$, $Nb_2C(OH)_2$, and $Ta_2C(OH)_2$ [61]. Early theoretical investigation reported several examples as shown in **Fig. 9a**, including $Sc_2CT_2$ (T=OH, F, O), $Ti_2CO_2$, $Zr_2CO_2$, and $Hf_2CO_2$ with band gap between 0.24 and 1.8 eV.

In particular, band gaps of 0.45, 1.03, and 1.8 eV, respectively, are reported for $Sc_2C(OH)_2$, $Sc_2CF_2$, and $Sc_2CF_2$, which demonstrate the impact of surface groups on the electronic characteristics of MXenes. Additionally, compared to $Sc_2CO_2$, the band structures of $Sc_2C(OH)_2$ and $Sc_2CF_2$ are more comparable. This is because O can accept two extra electrons, but F and OH groups only need one more electron to stabilize them. Except for $Sc_2C(OH)_2$, five of the six structures can be projected to have indirect band gaps based on their band topologies. Researchers have looked at the electrical structures of both bare and functionalized MXenes to better understand the mechanism that led to some MXenes becoming semiconducting after functionalization. The outcome demonstrates that, like MAX phases, all bare MXenes are metallic because the d orbital of the M element contributes all the Fermi energy **(Fig.9 f-g)**.

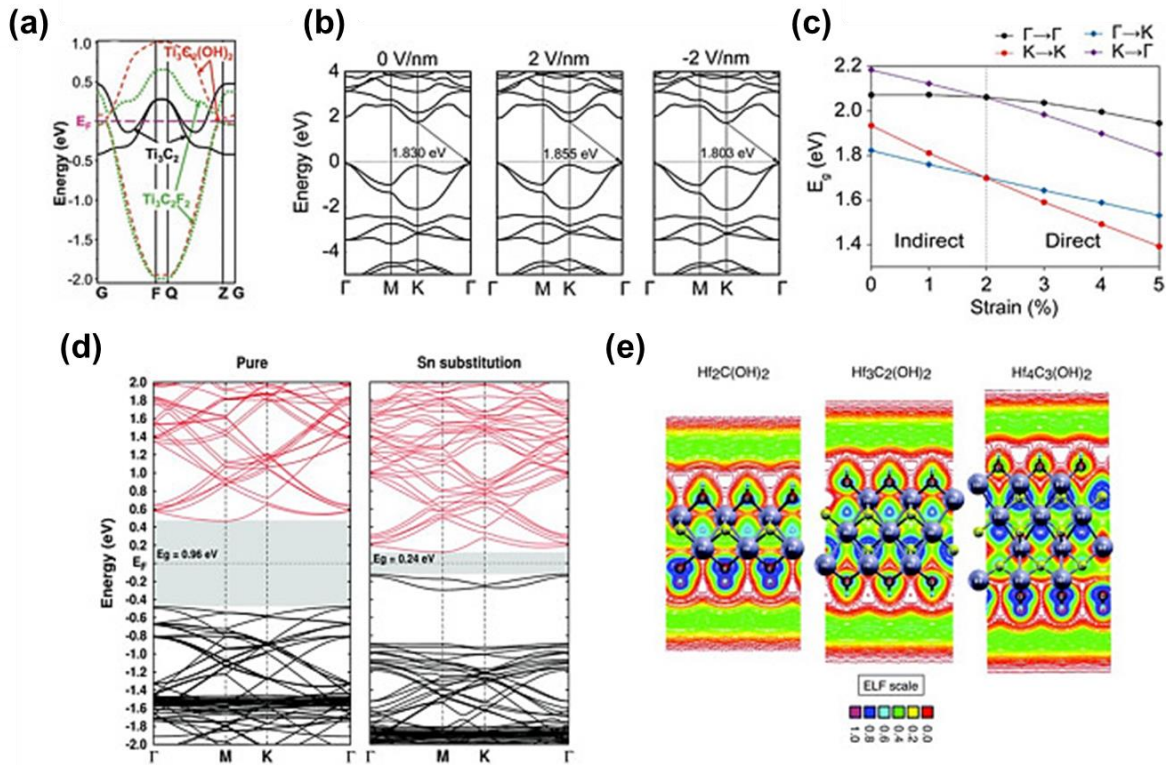

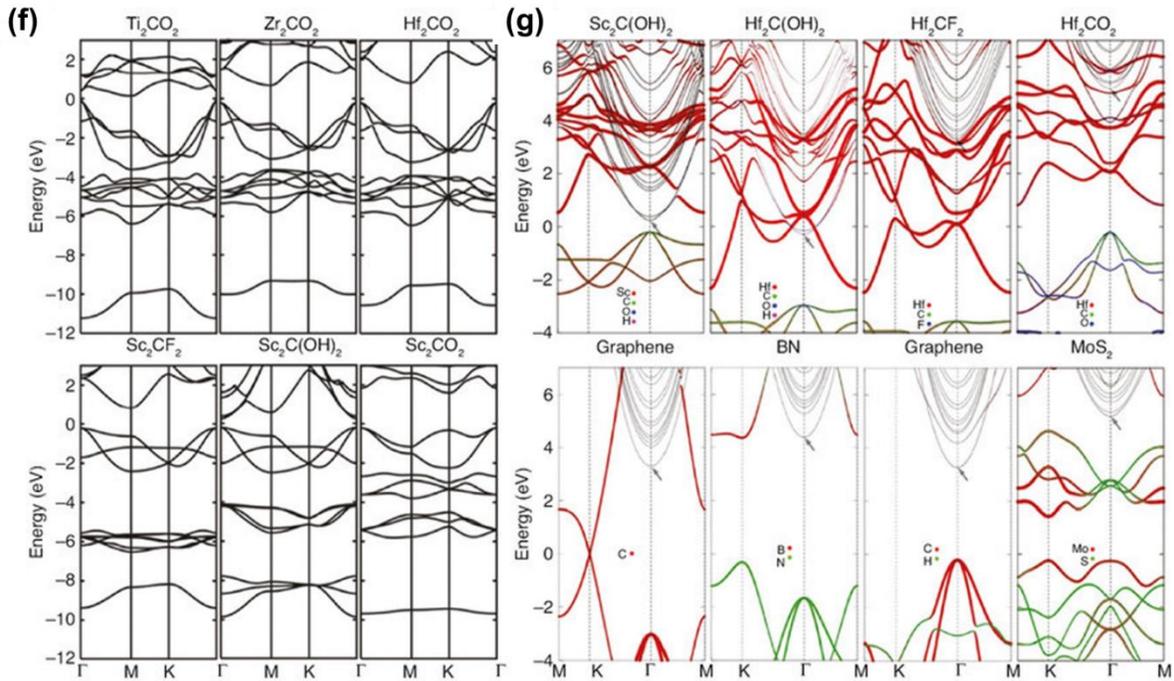

**Figure 9.** (a)-(e) The energy structure properties of Mxenes. (f)-(g) The band structures for six MXenes and the constituent atoms in the systems compared with graphene, boron nitride, and MoS$_2$ [61].

**Fig.10** provides a comprehensive overview of the electrical characteristics of functionalized MXenes. The vast MXene family can be categorized based on crystal structure, including the number of atomic layers and the presence of single metal, double metal, solid solutions, or ordered structures. The electronic properties of MXenes are strongly influenced by the specific transition metal element and surface functional groups. While the majority of MXenes exhibit metallic behavior, there are predictions of semiconducting MXenes with an ordered structure and the formula M$_2$CTx.

The subsequent sections delve into the surface functional groups, interlayer spacing, band structure, and work function of MXenes. MXenes have shown promise in various applications such as electrochemical energy storage, electromagnetic interference shielding, biomedical applications, and catalysis [55].

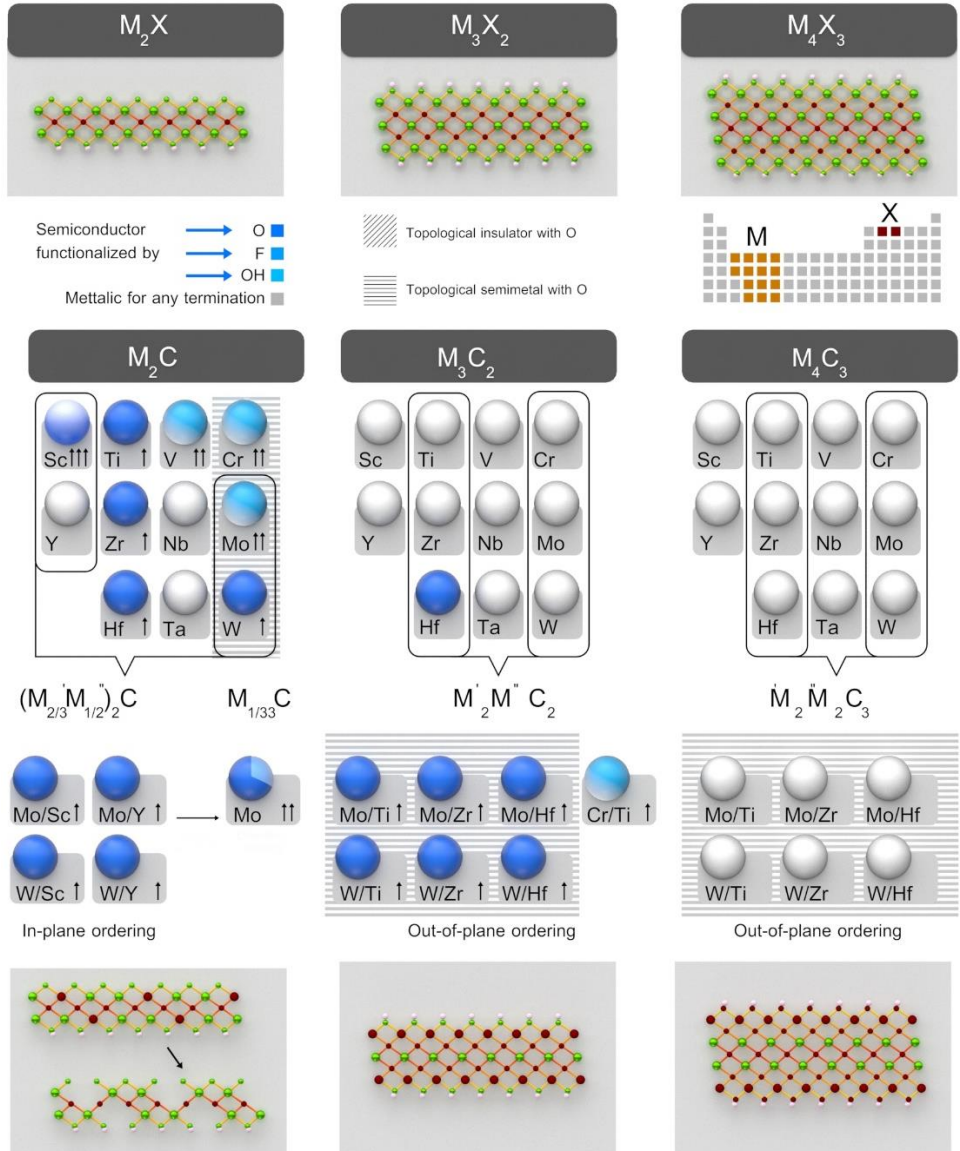

**Figure 11.** Diagrammatic representation of the electronic characteristics of the MXene family. The electronic characteristics of MXenes are categorized according to their structure, transition metal content, and functional group type. Metallic MXenes with any terminal groups are marked in grey, while semiconductor MXenes with the appropriate surface functional groups are highlighted with the relevant colors: blue, violet, and sky-blue colors represent oxygen, fluorine, and hydroxyl groups, respectively [55].

# V. Optical properties

Theoretical calculations of the dielectric function tensor have been employed to study optical properties such as absorption, reflection, and transmittance. Approximations like HSE06 have enabled investigations into the reflectivity, energy loss function, and absorption spectrum of bare $Ti_2C$, $Ti_2N$, $Ti_3C_2$, and $Ti_3N_2$. Plasmon energies have been determined for these structures, ranging from 10.00 to 11.62 eV. Depending on the circumstances, reflectivity can reach 100% or be less than 50% for electric fields parallel or perpendicular to the surface, respectively, suggesting the potential for transmitting electromagnetic waves at energies below 1 eV. The optical characteristics of functionalized $Ti_2C$ and $Ti_3C$, with surface terminations of -F, -OH, and -O, have also been theoretically explored alongside the bare structures. The study reveals that bare and O-functionalized MXenes (specifically $Ti_2C$ and $Ti_3C$) exhibit lower in-plane absorption coefficients than -F and -OH-functionalized ones across the infrared to ultraviolet (UV) light spectrum. It further suggests that -F- and -OH-terminated $Ti_2CT_2$ and $Ti_3C_2T_2$ should display complementary white colors. The linear optical properties (e.g., absorption, photoluminescence) and nonlinear optical properties (e.g., saturable absorption, nonlinear refractive index) of MXenes are highly influenced by energy structures, including bandgaps, direct/indirect bandgaps, and topological insulators, as well as the dispersion of the linear and nonlinear dielectric function ($\epsilon$) or refractive index (n = $\sqrt{(\epsilon\mu)}$) [61].

The optical response of materials is closely tied to their structural and electrical characteristics. Theoretical calculations allow us to distinguish between the real part (Re) and imaginary part (Im) of a material's dielectric constant as a function of frequency. In the case of a typical MXene like $Ti_3C_2T_x$, the termination (Tx) can significantly influence the structure, electrical properties, and ultimately the optical properties. The imaginary component of the dielectric constant is particularly affected by inter- and intraband electronic transitions, and these transitions contribute to the absorption spectra of the material. Hydroxylated and fluorinated terminations typically accept only one electron due to structural factors, resulting in similar behavior in the visible range. On the other hand, oxygen termination requires two electrons for stabilization. A recent study by Berdiyorov employed computational techniques to investigate the optical properties of $Ti_3C_2T_x$, highlighting the dependence on functionalization.

Functionalization resulted in a more than twofold reduction in the static dielectric function, as shown in **Fig 4a**. The oxidized sample exhibited higher absorption in the visible region, while surface fluorination led to lower absorption compared to pristine $Ti_3C_2$. The presence of oxygen atoms near the Fermi level significantly contributed to the total density of states of MXenes, leading to differences in optical performance between oxygen termination and hydroxylated/fluorinated terminations. Surface terminations increased reflectivity and absorption in the ultraviolet (UV) range compared to pristine $Ti_3C_2$. Oxygen termination enhanced reflectivity and absorption in the visible spectrum, while hydroxylated and fluorinated terminations showed lower reflectance and absorption than the pristine material. In certain long-term applications, the $Ti_3C_2T_x$ surface may undergo partial oxidation in a wet atmosphere.

In terms of optical absorption, the UV peak at 260 nm was larger, while the absorption peaks of $Ti_3C_2Tx$ at 780 nm and 325 nm were smaller. These findings demonstrate the possibility of tailoring the optical characteristics of MXenes by manipulating the surface termination for various applications. Moreover, when one carbon atom was replaced with a nitrogen atom in $Ti_3CNTx$, the typical absorption peak shifted to 670 nm instead of the 780 nm peak observed in $Ti_3C_2Tx$. Due to the dispersion of nitrogen's electrons in a lower energy band further from the Fermi level, $Ti_3CNT_x$ exhibited lower conductivity compared to $Ti_3C_2Tx$. The calculations for $Ti_3C_2$ employed the density functional theory [59].

## VI. Optoelectronic properties
### i) Plasmonic property

Plasmons are collective electron oscillations that occur when light interacts with matter. Surface plasmons (SP) are electron oscillations localized at the metal-dielectric interface, while bulk plasmons (BP) occur deeper within a system containing free carriers. Surface plasmons have found applications in various fields such as surface-enhanced Raman spectroscopy, photovoltaic devices, and optical and chemical sensors. MXenes, including $Ti_3C_2T_x$, are promising materials for plasmonic devices due to their strong metallic conductivity, tunable hydrophilicity, chemical stability, and scalable production.

The dielectric characteristics, including bulk and surface plasmons, of $Ti_3C_2T_x$ multilayers have been investigated using high-resolution electron energy-loss spectroscopy (EELS), scanning transmission electron microscopy (STEM), and ab initio calculations. In the mid-infrared range, the intensity of surface plasmons was found to be significantly stronger than that of bulk plasmons, as depicted in **Fig 12a**. Unlike other 2D materials like graphene, TMD, and h-BN, $Ti_3C_2T_x$ does not exhibit a clear blueshift in its bulk plasmon energy, which is attributed to its weaker interlayer interaction. The resonance of surface plasmons can be controlled by modifying the functional groups of $Ti_3C_2T_x$.

Further investigations using EELS and intensity mapping revealed the interband transition as well as transversal and longitudinal surface plasmon modes in mono- and multilayered $Ti_3C_2T_x$. These studies demonstrated that $Ti_3C_2T_x$ has independent polarizability due to weak interlayer coupling. By desorbing the F functional group at high temperatures, the energy of all surface plasmon modes shifted to blue. Arrays of $Ti_3C_2T_x$ nanodisks were also created to study localized surface plasmons, which exhibited strong resonances in the near-infrared range.

The dielectric permittivity of a $Ti_3C_2Tx$ film was experimentally measured, showing a crossover from dielectric to metallic behavior, indicated by Re dipping below zero at 1.07 μm. This is consistent with the plasmonic activity of the MXene film in the near- and mid-infrared regions. The introduction of a thin Au layer and an $Al_2O_3$ dielectric spacer layer further enhanced light absorption through the generation of gap surface plasmon resonance, as shown in **Fig. 12e** [60].

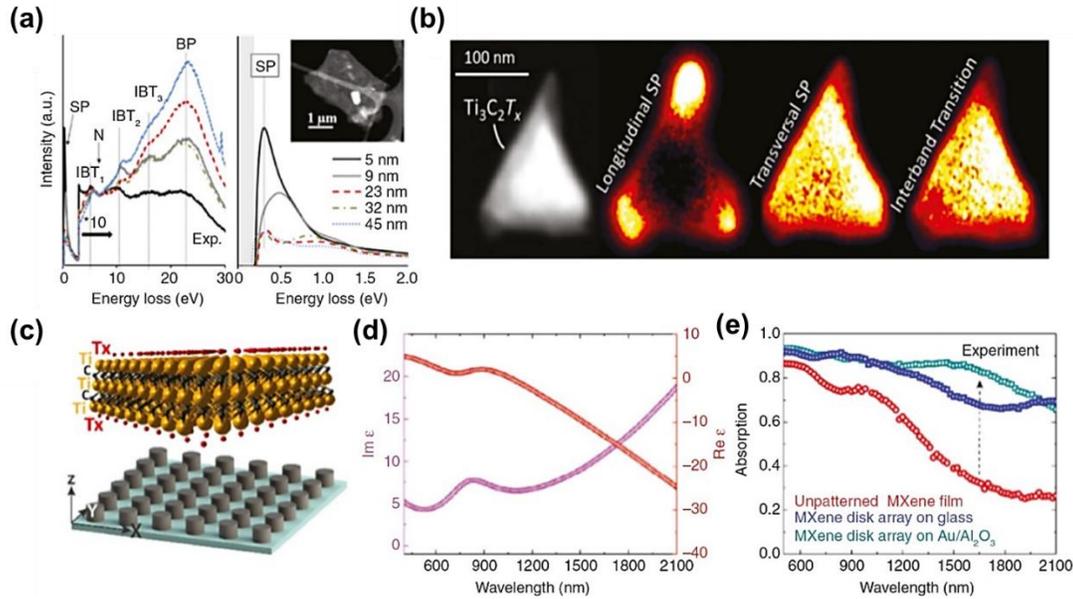

**Figure 12.** Demonstration of MXene surface plasmon. (a) Energy low-loss spectra of $Ti_3C_2(OH)_2$ with different materials and thickness profiles. Y-Scale magnified by an order of magnitude above 3 eV. The demonstration of surface plasmon is represented for thickness variations (right graph). (b) STEM-HAADF images of a triangular $Ti_3C_2T_x$ flake and corresponding EELS fitted intensity maps of SP and interband transition. (c) Graphical schematic representation of $Ti_3C_2T_x$ nanodisk array. (d) Experimentally measured permittivity (ε) of $Ti_3C_2T_x$ film with 400-nm thickness with real and imaginary refractive index plots. (e) Absorption spectra of unpatterned $Ti_3C_2T_x$ film and two types of nanodisk arrays were measured experimentally. Reproduced with permission [60]

## ii) Transparency and Conducting Property.

Transparent conductors with high conductivity, transparency, processability, mechanical characteristics, and flexibility are highly sought after for optoelectronic applications. Various deposition techniques such as spin-coating, dip-coating, spray-coating, and magnetron sputtering have been employed to fabricate transparent conductive MXene films. **Fig.13a** shows the linear optical transmittance of spray-coated $Ti_3C_2Tx$ films with controlled thickness, indicating a broad valley around 750 nm due to surface plasmon resonance. The transmittance of $Ti_3C_2Tx$ decreases with increasing thickness, but at a thickness of 70 nm, a transmittance of 43.8% was achieved. **Fig.12b** illustrates the transmittance of the $Ti_3C_2Tx$ film at 550 nm as a function of sheet resistance. The figure of merit (FoM), a commonly used metric, can be calculated based on electronic and optical conductivity. The maximum FoM obtained (0.74) is comparable to solution-processed graphene sheets.

To evaluate the performance of MXene films as flexible transparent electrodes, a $Ti_3C_2Tx$ film was deposited on a flexible polyester substrate. The inset of **Fig. 13b** demonstrates the change in sheet resistance (Rs) with bending radii. The Rs increased by approximately 15% under the reduced bending radius but returned to its initial value after the tension was released. This indicates that $Ti_3C_2Tx$ films possess good electromechanical properties, making them suitable for flexible optoelectronic applications.

Dillon et al. achieved the best FoM of 7.3 by spin-coating a $Ti_3C_2T_x$ film with an electronic conductivity of 6500 Scm$^{-1}$ and a transmittance of nearly 97%. By adjusting the concentration and spinning velocity during the spin-coating process, a $Ti_3C_2T_x$ film with an interconnected network and fewer boundaries exhibited a remarkably high conductivity of 9880 Scm$^{-1}$, resulting in the highest FoM of 19. This performance was comparable to the best carbon nanotube (CNT)-based films and solution-processed reduced graphene oxide (rGO). Furthermore, the conductance of a single $Ti_2CT_x$ nanosheet and the contact resistance between $Ti_2CT_x$ and chromium were evaluated using a four-probe approach. The I-V curve displayed in **Fig. 13c** showed an extracted conductance of 0.026 S and a low contact resistance of 0.243 mm, highlighting the potential of MXene as an electrode material for various optoelectronic device applications [60].

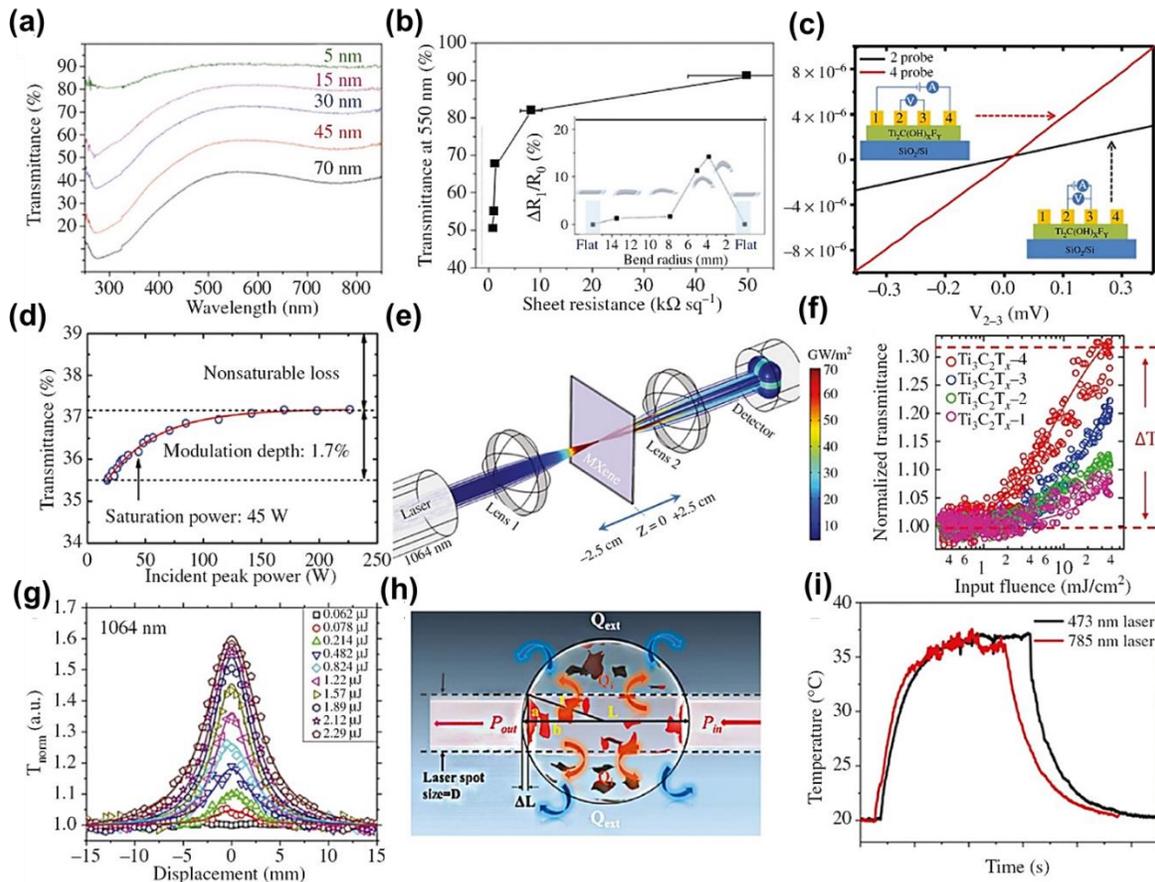

**Figure 13.** Optoelectronic properties of MXenes Optoelectronic properties of MXenes. (a) Transmission spectra of spray-coated $Ti_3C_2T_x$ films with variable thicknesses. (b) Transmittance was measured at 550 nm on sprayed $Ti_3C_2T_x$ films as a function of sheet resistance (kΩsq$^{-1}$). The inset shows a schematic and graphical plot representation of the bending characterizations of spray-coated $Ti_3C_2T_x$ films. (c) Four-probe measurements were performed to measure the conductance of a single $Ti_2CT_x$ nanosheet and contact resistance between $Ti_2CT_x$ and metal electrodes. The inset represents graphical schematics of a four-probe measurements cross-section. (d) The transmittance nonlinear energy-dependent curve of side-polished fiber with $Ti_3CNT_x$ monolayer film shows a modulation depth of 1.7%. (e) Graphical representation of the setup of open aperture Z-scan method. (f) Normalized nonlinear transmittance of $Ti_3C_2T_x$ films (-1, -2, -3, -4) as a function of input fluence. (g) Open aperture Z-scan characterizations of position-dependent nonlinear transmittance of $Ti_3C_2T_x$. (h) Graphical schematic of the droplet of $Ti_3C_2T_x$ solution with laser

illumination. (i) Time-resolved temperature profile of $Ti_3C_2T_x$ droplet under two separate laser irradiations. Referred to [60].

### iii) Nonlinear optics property

Nonlinear optics explores the interaction between the optical field and electrons and phonons when materials exhibit nonlinear responses to applied electromagnetic fields. Saturable absorptions (SA), optical rectification, Kerr effects, and harmonic generation are examples of phenomena that arise due to nonlinearity and find wide applications in photonics, optics, and flexible electronics. $Ti_3C_2Tx$ demonstrates a fundamental metallic behavior, allowing it to span a wide range of frequencies, including the near-infrared (NIR) and far-infrared (FIR) regions.

The nonlinear transmission curve of $Ti_3C_2Tx$, as depicted in **Fig. 13d**, exhibits unique characteristics. The open aperture Z-scan technique, illustrated in Fig.13E, was employed to study the nonlinear response. The difference in normalized transmittance between high and low illumination intensities, known as the modulation depth (T), is observed. **Fig.13f** presents the nonlinear transmission as a function of laser intensity. The saturable absorption behavior observed in $Ti_3C_2Tx$ films of varying thicknesses is attributed to an increase in ground state absorption at 1064 nm induced by plasmons. The thickness was found to be adjustable to control the modulation depth and saturation fluence. $Ti_3C_2Tx$ exhibited a higher threshold for light-induced damage compared to other two-dimensional materials. The nonlinear optical response of $Ti_3C_2Tx$ was investigated across a broad spectral range from 800 nm to 1800 nm. It was observed that the effective nonlinear absorption coefficient decreases as the pulse energy is changed, indicating that one-photon processes dominate the nonlinear absorption at lower pulse energies, while other processes occur at higher energies **(Fig.13g)** [60].

### iv) Photothermal conversion property

MXenes such as $Ti_3C_2$, $Nb_2C$, and $Ta_4C_3$ have shown great potential in photothermal therapy due to their 2D shape, strong absorption in the near-infrared (NIR) range, and high conversion efficiency. The localized surface plasmon, believed to be responsible for their NIR light absorption and photothermal conversion, plays a crucial role in their efficacy. Recent studies have focused on investigating the photothermal conversion property of $Ti_3C_2Tx$.

In these studies, a droplet of aqueous solution containing $Ti_3C_2Tx$ was irradiated with a laser of a specific wavelength. The temperature of the droplet was monitored in real-time using an infrared (IR) camera. As shown in Fig.13H, the temperature of the droplet increased rapidly upon laser activation and sharply decreased when the laser was turned off, as depicted in the temperature profile shown in **Fig.13i**. Remarkably, the internal photothermal conversion efficiency reached up to 100% [60].

## VII Magnetic properties

The two-dimensional layers of MXenes exhibit a wide range of magnetic properties. Among MXenes, 2D $Ti_3C_2$ and $Ti_3N_2$ are predicted to be antiferromagnetic, while 2D $Cr_2C$, $Cr_2N$, $Ta_3C_2$, and $Cr_3C_2$ are expected to be ferromagnetic when exfoliated from their MAX phases. Ti2C and Ti2N show nearly half-metallic ferromagnetism. The monolayer Mn2C is an antiferromagnet with a high Neel temperature of 720 K, but functionalization with F, Cl, and OH can induce ferromagnetism with a high Curie temperature (520 K). Although some bare MXenes are predicted to be ferromagnetic, experimentally synthesized MXenes typically have F, O, or OH terminations. Pure MXenes such as Cr2C are challenging to synthesize, but recent research has shown that partial surface removal can enhance the electrical conductivity of MXenes.

Functionalized $Cr_2C$ and $Cr_2N$ MXenes (F, O, H, Cl, and OH) exhibit magnetic properties, but some display ferromagnetic-to-antiferromagnetic (FM-AFM) transitions. The magnetic characteristics of these MXenes arise from the d-orbitals of the Cr atoms. The significant energy difference between the FM and AFM configurations of Cr-based MXenes suggests that their magnetic properties can be maintained up to room temperature. Non-terminated $Cr_2C$ with a half-metallic gap of 2.85 eV can exhibit ferromagnetism, while surface terminations (F, OH, H, or Cl) can induce FM-AFM and associated metal-to-insulator transitions by localizing the d electrons. Ferrimagnetic half-metals with 100% spin polarization of electrons around the Fermi level hold promise for spintronic applications. Oxygen-passivated $Cr_2NO_2$ shows a ferromagnetic ground state with a significant half-metallic gap, while $Cr_2N$ MXenes functionalized with F and OH exhibit antiferromagnetic properties. $Cr_3C_2$ is a highly resistant ferromagnetic monolayer with a total magnetic moment of 3.9 μB per formula unit. It behaves as a half-metallic ferromagnet with a gap of 1.2 eV, and this behavior can be sustained under strain. Previous investigations have explored the magnetic properties of Cr-based carbides ($Cr_2M_2C_3T_2$; M = Ti, V, Nb, and Ta; T = OH, O, and F), showing that $Cr_2Ti_2C_3O_2$ and $Cr_2V_2C_3O_2$ are energetically favorable ferromagnetic arrangements, while other systems exhibit FM-AFM phase transitions with extensile strain variations **(Fig. 14f)**.

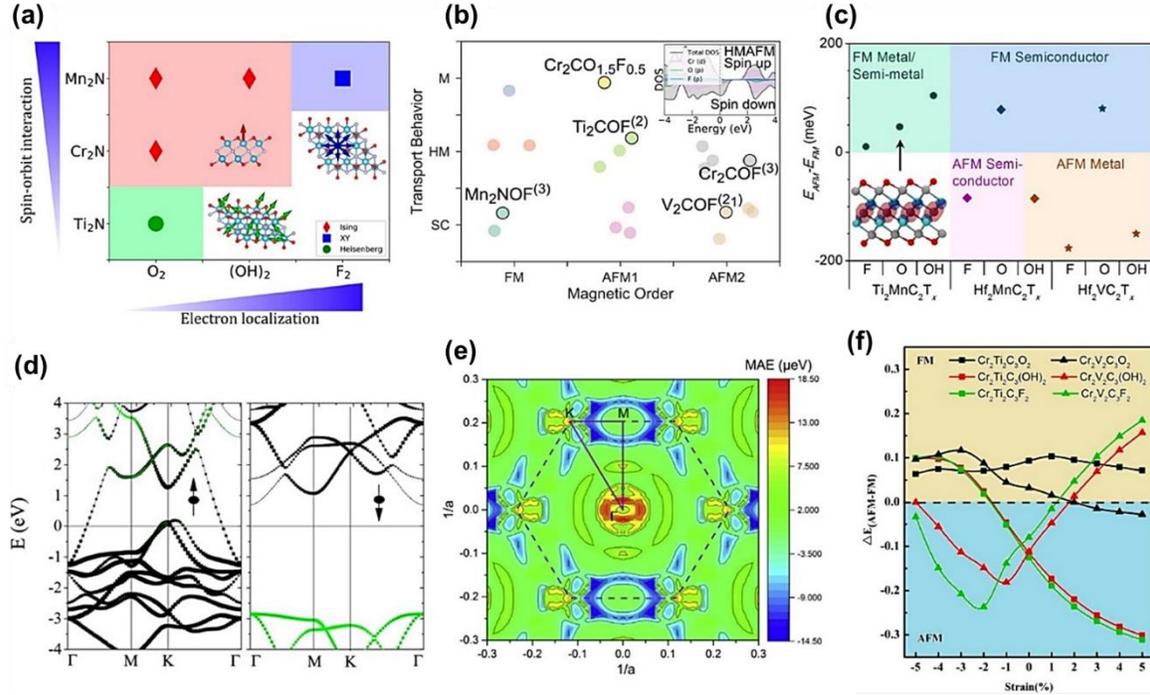

**Figure 14.** (a) Representation of available spin states of $M_2N$ (M=Ti, Cr, Mn) based on spin-orbit coupling and electron localization effect. (b) Scatter plot showing the diversity of transport and magnetic behavior observed in Janus MXenes. (c) Exchange energy and magnetic states in mixed Ti and Hf-based MXenes with various surface termination. (d) Band structure of different spin states for half-metallic $Cr_2C$ MXene. The black and green squares indicate the weights of the Cr d and C p orbitals, respectively. (e) The large negative magnetic anisotropy energy around the sides of the hexagonal Brillouin zone of $Fe_2C$ MXene. (f) Magnetic properties of $Cr_2M_2C_3T_2$ (M=Ti and V, T=OH, O, and F) at different extensile strains. Referred to [61].

## VIII. Topological properties

While most MXenes exhibit metallic or semiconducting behavior, certain functionalized MXenes have been predicted to be two-dimensional topological insulators (TI). In these materials, the presence of heavy elements leads to a relativistic phenomenon known as spin-orbit coupling (SOC), which significantly affects the electronic structure and enables dissipation-free transport along edge states, making them suitable for low-power electronic devices. The observed topological insulating MXenes often contain Mo, W, Ti, Zr, and Hf atoms. For example, O-functionalized $M_2CO_2$ (M = W, Mo, and Cr) MXenes exhibit a 2D TI phase with significant SOC-induced bandgaps ranging from 0.194 to 0.472 eV when calculated using the GGA (HSE06) functional **(Fig. 14a)**. In the $Ti_2CF_2$ phase, multiple Dirac cones and substantial spin-orbit splitting are observed. Importantly, the 2D band structure of MXenes remains intact even in multilayer phases or under external perturbations [61].

## IX. Electronic work functions

The electronic work function (WF) of MXenes represents the energy required to remove an electron from the surface, and it is influenced by electronic redistribution and induced surface dipoles from functional groups. The altered work function can be expressed as WF = (e/0)P, where P denotes the altered total dipole moment. In general, MXenes exhibit the following work function trend: O$^-$>F$^-$>bare>OH$^-$ functionalized MXene **(Fig. 15a)**. Among various terminated MXenes (Sc$_2$X, Ti$_2$X, V$_2$X, Cr$_2$X, Zr$_2$X, Nb$_2$X, Mo$_2$X, Hf$_2$X, and Ta$_2$X, where X = C or N), the lowest anticipated work function is around 1.2 eV **(Fig. 8b)** due to the hybridization of metal d orbitals and O 2p orbitals. The methoxylated niobium carbide Nb$_3$C$_2$(OCH$_3$)$_2$ MXene is predicted to have an exceptionally low work function of 0.9 eV. This ultra-low work function makes MXenes promising for thermionic and field-emitter cathodes. The work function of Ti$_3$C$_2$T$_x$ MXene can be adjusted between 3.9 and 4.8 eV by annealing at different temperatures, which alters the surface termination moieties **(Fig.15c)**. However, the ultra-low work function of Ti$_3$C$_2$T$_x$ has not been experimentally verified, and a Kelvin probe measurement on a terminated Ti$_3$C$_2$ MXene thin film determined a work function of 5.28 eV [61].

In metal-semiconductor junctions, the Mott-Schottky rule is often used to predict or explain the contact behavior. Depending on the difference between the work function of the metal and the electron affinity or ionization energy of the semiconductor, Ohmic contact or Schottky junctions with specific barrier heights can be established. However, experimental results often deviate from the fundamental theory due to interfacial defects and the Fermi-level pinning effect. High-energy metal deposition techniques like e-beam evaporation can generate gap states and cause damage to the underlying material. Recent studies have shown that van der Waals metal-semiconductor contacts can be created using low-energy metal deposition techniques, resulting in distinct operating characteristics for transistors. Therefore, low-energy metal deposition techniques are crucial, and MXenes processed via solution methods hold promise as a useful option [55].

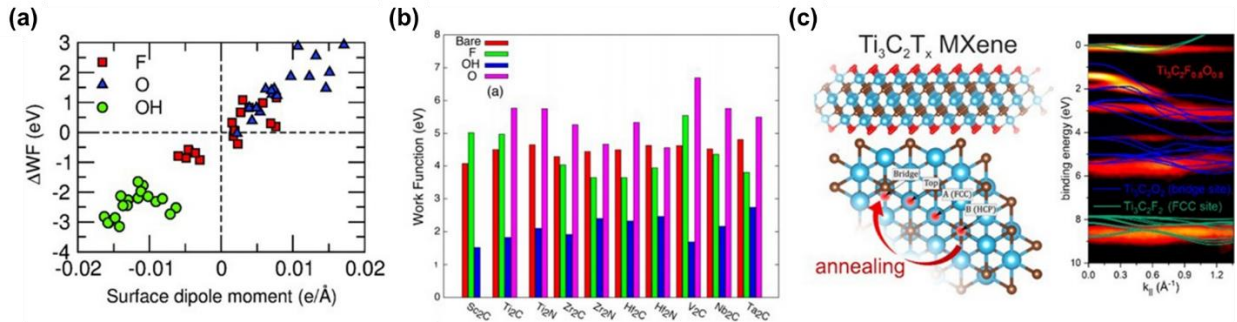

**Figure 15.** (a) Work function changes as a function of surface termination and their corresponding surface dipole moment. (b) The work functions of M2C and M'2N (M = Sc, Ti, Zr, Hf, V, Nb, Ta; M' = Ti, Zr, Hf) functionalized with F, OH, and O termination. (c) Annealing treatment of Ti3C2Tx MXene leads to the modification of surface moieties and work function. The right figure shows the measured ARPES curvature spectrum of Ti3C2F0.8O0.8 (yellow-red), compared to two calculated energy structures of Ti$_3$C$_2$O$_2$ (blue) and Ti$_3$C$_2$F$_2$ (green) [61].

## X. MXene Based Heterostructures and Interfaces

MXenes and other 2D materials provide a platform for constructing van der Waals (vdW) heterostructures with intriguing electrical properties **(Fig. 16a)**. These heterostructures can be engineered and tuned through proximity effects, charge transfer, and interface-induced strain, enabling manipulation of their electronic properties. The creation of heterostructures impacts the electronic structure of MXenes, offering opportunities for studying their interactions with a wide range of materials, including transition metal dichalcogenides, even with minimal lattice misfit. Heterostructures such as $MoS_2/Ti_2C$ exhibit strong chemical bonding due to unsaturated surface states, like the bonding observed in graphene/$Ti_2C$. On the other hand, $MoS_2/Ti_2CF_2$ and $MoS_2/Ti_2C(OH)_2$ interfaces demonstrate non-covalent vdW-type interactions. The pronounced metallic character of $Ti_2C$ transforms $MoS_2$ from a semiconductor to a metallic type, while maintaining its semiconducting properties in other heterostructures. Hetero-bilayers of TMD/$Sc_2CF_2$ can be converted into type-II indirect bandgap semiconductors with energy gaps ranging from 0.13 to 1.18 eV. Similarly, $MoS_2/TM_2CO_2$ composites (where TM is Ti, Zr, or Hf) exhibit type-II semiconducting characteristics. By combining multiple MXene layers, different types (I, II, and III) of heterostructures can be created, as demonstrated with 2D $Sc_2C$ ($Sc_2CF_2$, $Sc_2C(OH)_2$, and $Sc_2CO_2$) systems. Type-I heterostructures result from stacking F and O-functionalized MXenes, while F/OH and O/OH systems form type-III heterostructures. Equibial strain can be employed to engineer type-II hetero-bilayers, as shown in **Fig. 15c**. Machine learning techniques have been applied to accurately predict band locations in various MXenes. Additionally, transport calculations have identified $MoS_2$ and $Ta_2CF_2/Ta_2C(OH)_2$ as potential electrode materials with Ohmic contacts. The n-type Schottky barrier in $MoS_2/Ti_2CF_2$ and $MoS_2/Ti_2C(OH)_2$ heterojunctions is measured to be 0.85 eV and 0.26 eV, respectively. Through surface functionalization and applied electric fields, the Schottky barriers in semiconductor/metal heterostructures can be controlled [61].

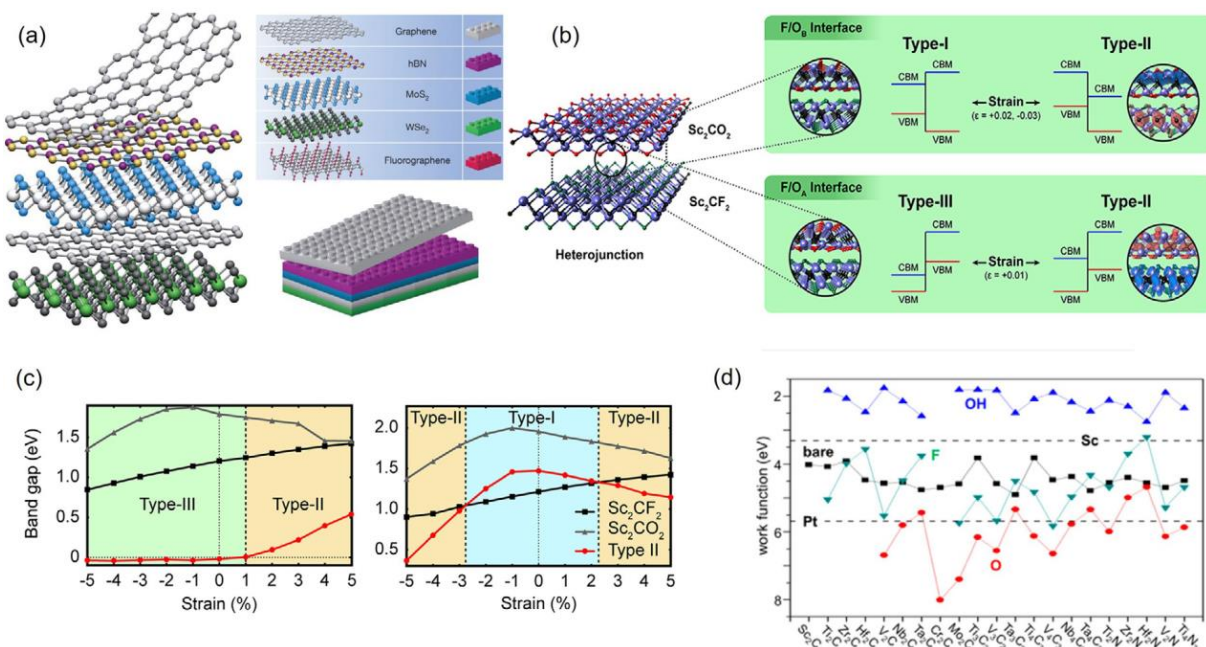

**Figure 16.** Schematic illustration of multilayer heterostructure formation [61].

## Applications of MXenes

## I. Energy storage (Batteries)

2D MXene sheets exhibit high electrical conductivity and hydrophilicity, making them potential candidates for electrochemical ion batteries. The conductivity of MXenes can range from metallic-like to semiconducting or insulating, depending on the characteristics of the M elements and terminal functional groups. Previous research has shown that MXenes based on Ti and Mo with fewer defects and good flake contact demonstrate high conductivity, such as $Ti_2CT_z$ and $Ti_2C_2T_z$. The interlayer spacing also influences conductivity. Single-layer MXenes or MXenes without terminations are theoretically expected to exhibit metallic conductivity.

The large interlayer spacing in MXenes enables ion transport between the layers, making them promising electrode materials for secondary ion batteries, including Li-ion, Na-ion, K-ion, and Al-ion batteries, due to their high conductivity. Among these, MXenes have been extensively studied for Li-ion batteries (LIBs). Monolayer $Ti_3C_2T_z$, for example, exhibits a capacity of approximately 150 mAh/g at 260 mA/g in LIBs, compared to around 380 mAh/g for carbon-based electrodes. However, during the charge/discharge process, MXene sheets may stack or collapse, leading to fewer active sites and hindered ion transport between layers, particularly in the initial cycle. Strategies such as using delaminated or intercalated MXene flakes and stable interlayer structures have been explored to enhance the capacity and stability of MXene-based electrodes. For instance, the capacity of LIBs was quadrupled by employing carbon nanotube (CNT) spacers and porous $Ti_3C_2T_z$ flakes to prevent restacking. Intercalating Sn+ or other cations between MXene sheets is another approach to stabilize the interlayer structure. Hybridization of MXenes with other materials has also been investigated to enhance battery performance, as demonstrated in Table 1. Notably, the hybrid material $MoS_2/Ti_3C_2$-MXene displayed a high capacity of approximately 1200 mAh/g at 200 mA/g. MXene-based electrodes in Li-S batteries and other ion batteries also show promising potential. Given their high theoretical capacity compared to graphene, researchers are actively exploring MXenes in this field [4].

| Batteries | Materials | First C.E. | Initial discharge/charge capacities (mAh/g) | Cyclability (mAh/g) |
|---|---|---|---|---|
| LIB | MoS2/Ti2C3-MXene@C | 69.1% | 1210/1750 @ 200 mA/g | 1200 @ 100 mA/g after 3000 cycles |
| SIB | Ti2C3Tx nanodot-P composite | 66% | 600/909 @ 100 mA/g | 400 @ 400 mA/g after 150 cycles |
| Li-S | S/L-Ti3C2 | 99.8% | 1288/1291 @ 200 mA/g | 970 @ 200 mA/g after 100 cycles |

Table 1: Comparison of performance of batteries with MXene-based electrodes [58].

## II. Supercapacitors (Energy storage)

Supercapacitors are a type of energy storage device known for their high energy density and fast charge/discharge capabilities. There are two main types of supercapacitors: electrical double-layer capacitors (EDLC) and pseudo-capacitors, which differ in their energy storage mechanisms. EDLCs store charges by creating an electrical double layer on the electrode surface, while pseudo-capacitors store charges through redox processes on the electrode surfaces. Pseudo capacitors generally offer higher stability and higher charge storage capacity than EDLCs.

MXenes, with their high electrical conductivity and large specific surface area, have emerged as promising electrode materials for supercapacitors. MXene electrodes primarily retain charges through the intercalation of polar organic molecules or cations in redox reactions. Additionally, MXene electrodes exhibit significant capacitance even at high scan rates, where the rate of Faradic charge transfer becomes the limiting step. This indicates that the MXene surface can store electrostatic charges effectively. MXenes with high density demonstrate high volumetric capacitance and excellent cyclability compared to carbon-based materials.

$Ti_3C_2T_x$ is one of the most extensively studied MXenes for supercapacitor applications. It exhibits a volumetric capacitance of 300 to 400 F/cm3 in basic and neutral electrolytes, which is comparable to activated graphene-based electrodes. In an acidic electrolyte (1 M $H_2SO_4$), a rolled $Ti_3C_2T_x$ clay electrode has shown a remarkable volumetric capacitance approaching 900 F/cm$^3$. MXenes' exceptional performance in supercapacitors holds great potential for high-performance energy storage applications. Several factors contribute to the remarkable performance of MXene-based supercapacitors. Firstly, MXenes offer a larger number of electrochemically active sites, particularly for small cations, which enhances the charge storage capacity. During the production process, the intercalated Li+ ions hinder the restacking of MXene sheets, leading to improved redox processes.

$Ti_3C_2T_x$-based supercapacitors exhibit excellent cyclability, maintaining their capacitance even after 10,000 cycles. Surface modification of MXenes can further enhance their performance by impacting the terminal groups. For instance, removing or replacing the -F groups with -O-containing groups significantly increases the capacitance. Modified $Ti_3C_2T_x$ with fewer terminal groups and more electrochemically active sites has shown a 211% higher capacitance compared to pure $Ti_3C_2T_x$. Replacing -F groups with -OH groups can double the capacitance, and in acidic electrolytes, the improvement can be as high as seven times the initial value.

Other MXene materials, such as $Ti_3C_2T_x$ and $Mo_2CT_x$, both in their pristine form and in modified structures (such as with polypyrrole and PEDOTS), also demonstrate excellent volumetric capacitance. The choice of electrolyte and counter electrode also plays a significant role in optimizing the performance of MXene-based supercapacitors. By carefully selecting MXenes, electrolytes, and counter electrodes, further improvements can be achieved in the performance of MXene-based supercapacitors.

## III. Transparent conductive thin films in Optoelectronics

$Ti_3C_2T_x$ MXene thin films exhibit a combination of high transmittance and electrical conductivity, making them promising for optoelectronic applications. The characteristics of the films can be influenced by the surface termination groups and microstructures. It has been observed that MXene films produced using aqueous suspensions through spray coating exhibit higher transmittance compared to those produced using ethanol-based suspensions. Intercalation of TMAOH has shown the potential to increase transmittance by more than 20%.

Spin-coated $Ti_3C_2T_x$ MXene thin films with a thickness of 4 nm, characterized by horizontally oriented large flakes, demonstrate excellent transparency of 93% and high electrical conductivity of 5736 S/cm. MXenes possess exceptional mechanical properties, high electrical conductivity, relatively high transparency, 2D morphology, and processability, making them ideal for creating transparent conductive thin films. Extensive theoretical and experimental studies have been conducted on the optical properties of MXenes, with particular focus on $Ti_3C_2T_x$. Various techniques have been developed to produce transparent MXene films, including the etching of $Ti_3AlC_2$ precursor to creating $Ti_3C_2T_x$-IC films on a sapphire substrate. These films intercalated with $NH_4HF_2$, can achieve over 90% transmittance, surpassing regular MXene films with transmittance values of around 70% to 30%. Transmittance spectra reveal the difference between MAX and MXenes, with MXenes showing an absorption peak around 300 nm **(Fig.17)**.

Solution processing methods, such as spin casting, have been employed to produce MXene thin films with nanometer-scale thickness and impressive electrical conductivity of up to 6600 S/cm while maintaining high transparency of 97%. Vacuum annealing has been found to significantly improve the conductivity of MXene films without compromising transparency. The flexibility of MXene films on a bendable substrate slightly reduces their conductivity compared to glass, but the bend cycle has minimal impact on conductivity. Among the solution processing techniques, spin casting has shown superior performance in producing MXene films. However, it requires relatively high concentrations of MXene dispersion (>5 mg/ml) to achieve noticeable film thickness. The substrate must be flat to achieve the best conductivity due to the limitations of the spin chamber. Spray coating offers a simple and controllable alternative for thin film deposition, allowing deposition over large areas with low solution concentrations (0.5-3 mg/ml). However, the resulting films often exhibit a rough surface with granular boundaries, posing challenges for industrial applications.[4]

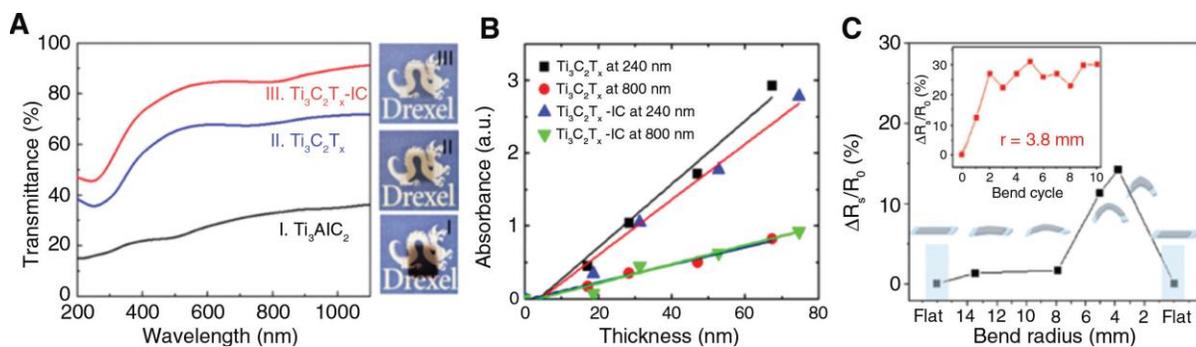

**Figure 17.** The change in film resistance upon bending, as well as the light absorbance spectra of various MXenes [58].

## IV. Sensors

The above analysis highlights the significant features of MXenes in the fields of thermoelectrics and optoelectronics, positioning them as important 2D materials alongside counterparts like $MoS_2$ and black phosphorus. Consequently, research teams have explored the use of MXenes as sensor materials, with promising results. One particularly crucial application is in pressure sensors, where the strong correlation between the cross-plane conductivity and interlayer spacing of MXenes proves advantageous. MXenes exhibit excellent sensitivity to mechanical strain and external pressure, surpassing carbon-based materials, metal wires, and MoS2. For instance, a porous 3D MXene sponge created through dip coating has demonstrated the ability to detect mechanical pressures as low as 9 Pa. Such high sensitivity makes MXenes suitable for the development of pressure sensors for human-machine interfaces. By sandwiching conductive MXene nanosheets ($Ti_3C_2T_x$) between two polylactic acid sheets, a pressure sensor capable of detecting a broad range of up to 30 kPa has been realized. This technology enables continuous monitoring of human activity, such as tracking muscle and joint movements, which holds great potential in medical rehabilitation [58].

In addition to pressure sensors, MXenes have also shown promise in electrochemical biosensors for detecting a range of analytes, including volatile organic chemicals, water vapor, dopamine, illness biomarkers, environmental toxins, and more. Various MXene-based sensors have been developed to measure mechanical strain, gases, and biomolecules. MXene films, composed of flexible and electrically conductive nanosheets, maintain their functionality even when subjected to mechanical deformations. The horizontally stacked structure of MXenes facilitates the formation of diverse electrical channels through Coulombic charge interactions between the negatively charged MXene surface and positively charged intercalants or dipolar water molecules. These characteristics make MXenes well-suited for wearable sensor applications.

| Electrode | Analyte | Detection method | Detection limit | Detection range |
|---|---|---|---|---|
| **Ti₃C₂Tₓ-MXene** | Carbendazim | DPV | 103 nm | 50 nm–100 μm |

| Hb/Ti$_3$C$_2$eGO/gold foil | H$_2$OS$_2$ | Amperometry | 1.95 mm | 2 mm–1 mm |
|---|---|---|---|---|
| Ti$_3$C$_2$T$_x$/GCE | H$_2$O$_2$ | Chronoamperometry | 0.7 nm | - |
| Anti-CEA/f-Ti$_3$C$_2$/GCE | CEA | CV | 18 fg/ml | 100 fg/ml$^{-2}$ µg/ml |
| MXenes/Apt2/exosomes | MCF-7 | ECL | 125 particles/µl | 5x10$^2$-5x10$^6$ |
| Urease/MB-MXene/SPE | UA, urea | SWV | 5, 0.02 µm | 30–500, 0.1–3.0 µm |
| Alke Ti$_3$C$_2$/GCE | Cd(II), Pb(II), Cu(II), Hg(II) | SWASV | 0.098, 0.041, 0.032, 0.130 µm | 0.1–1.5 µm |

Table 2: Comparison of various electrodes based on the types of analytes and detection characteristics [58].

Delaminated Ti$_3$C$_2$T$_x$ MXene has been employed as an electrode modification for the detection of the fungicide carbendazim, targeting environmental pollutants. Additionally, the modified Ti$_3$C$_2$Tx/GCE electrode has been utilized to measure the concentrations of heavy metallic ions, including Cd, Pb, Cu, and Hg. The presence of H$_2$O$_2$, a reactive oxygen species with significant biological functions, is frequently assessed through oxidase reactions. H$_2$O$_2$ detection can be achieved enzymatically or nonenzymatically. The Ti$_3$C$_2$Tx/GCE Hb/Ti$^+$C$_2$eGO/gold foil electrode MXene sensor demonstrates the capability for direct electron transfer between the enzyme and electrode. This enables the quantification of H$_2$O$_2$ concentrations due to the electrode's high surface area and favorable electronic properties [4].

## V. Biomedical applications

MXenes have a promising future in the field of biomedical applications, encompassing biosensors, drug delivery systems, photothermal therapy (PTT) for cancer treatment, bioimaging, and more. This is attributed to their hydrophilicity, large specific surface areas, excellent biocompatibility, biodegradability, high near-infrared (NIR) absorption, and efficient light-to-heat conversion capability. These qualities make MXenes highly suitable for diverse biomedical applications. In PTT, MXenes act as photothermal agents, harnessing the heat generated by NIR laser irradiation to destroy cancer cells. Extensive research has demonstrated the effectiveness of MXenes in eradicating tumors in vivo and damaging cancer cells in vitro. For example, dual therapy nanoplatforms based on tin sulfide nanosheets, when combined with chemotherapy, have shown remarkable efficacy in eliminating cancer cells in mice without causing harm to healthy organs. The hydrophilic nature and large surface areas of MXenes also make them ideal for drug delivery applications. Previous studies have demonstrated that MXenes, such as Ti$_3$C$_2$, can efficiently transport drugs with a high loading capacity (e.g., 211.8%) for chemotherapy. This indicates the potential for synergistic therapy by combining PTT and chemotherapy using MXenes. Moreover, MXenes have shown promise in various other biomedical applications, including biosensing, bioimaging, theranostics, and antibacterial properties.

Surface modifications enable deeper exploration of the potential of MXenes. For instance, PEGylation of $Ti_3C_2$ enhances its stability in physiological solutions, while termination with an Al oxoanion increases its photothermal effect. Combining MXenes with MnOx or GdW10-based polyoxometalates or mesoporous silica nanoparticles offers versatile applications in therapeutics, diagnostic imaging, magnetic resonance, computed tomography, and more.

To ensure their safety, the biocompatibility of MXenes has been extensively studied. Previous investigations have demonstrated the good cytocompatibility of $Ti_3AlC_2$, $Ti_3SiC_2$, and $Ti_2AlN$. Studies have also shown that $Ti_3C_2$ can be easily eliminated from test subjects through urine and feces. Photodegradability of tin sulfide nanosheets has been observed after prolonged laser irradiation. MXene quantum dots (QDs) created using a fluorine-free technique exhibit excellent biocompatibility. However, further research is needed to fully understand the toxicity and biocompatibility of MXenes, particularly in vivo, before their widespread use in biomedical applications. MXenes have shown great potential in various biological processes, including photothermal therapy, drug delivery nanoplatforms, biosensing, bioimaging, theranostics, and antibacterial applications. Future investigations can focus on surface modifications and further exploration of their toxicity. Figure 17 provides an overview of MXene applications in biomedicine [58].

## VI. Optoelectronic device applications

### i) 2D Vanadium Carbide-based Photodetectors.

Vanadium carbide ($V_2C$), a newly developed MXene, has shown great potential as an excellent nonlinear optical material. In this study, we report the synthesis of few-layer $V_2CT_x$ nanosheets using the liquid phase exfoliation (LPE) method. MXenes, a family of 2D transition metal carbides, nitrides, and carbonitrides, have demonstrated fascinating characteristics including adjustable band gap, high optical transparency, super high conductivity, and excellent environmental stability [63], [64]. These materials have also exhibited promising performance in various applications such as electrochemical capacitors, chemical catalysts, biosensors, superconductivity, and energy storage. The unique properties of functional groups associated with MXenes provide them with rich characteristics, and the ability to independently adjust monolayer thickness regardless of chemical composition and bonding type adds to the allure of MXenes [46]. While the exploration of MXenes is still in its infancy, the investigation of different types of MXenes and their properties is crucial. $V_2CT_x$, as a novel MXene, has garnered significant interest due to its exceptional electron and ionic conductivity, as well as long-term stability under ambient conditions, surpassing even $Ti_{n+1}C_nT_x$ [65].

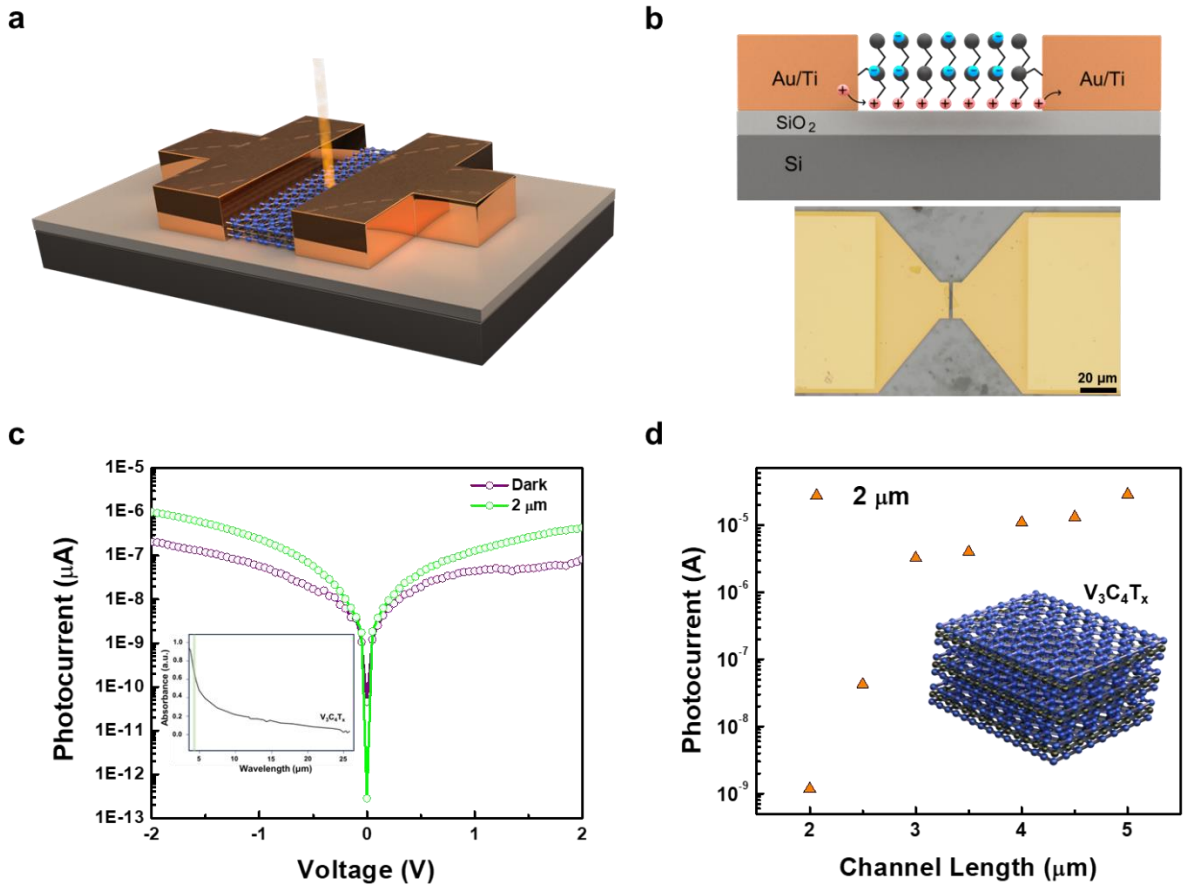

**Figure 18.** 2D Vanadium Carbide (MXene) mid-IR photodetectors. a) Schematic representation of the photodetector device. b) Cross-sectional view of the device representing the material layer and Au/Ti Contacts. The microscope image shows the passivation PMMA layer on top and the etched contact area on the electrical contact pads. c) Current-voltage (I-V) characteristics of the photodetector can be observed for dark current and photocurrent response at 2 um laser illumination from the top. Inset shows the FTIR measurement of the vanadium carbide flakes demonstrating the photo absorption response of the material. d) Variation in photocurrent as a function of channel length shows the increase in responsivity per unit area [46], [63]-[65].

In this study, we demonstrate a photodetector device utilizing vanadium carbide ($V_2C$) for mid-infrared (IR) sensing and detection. The fabrication process involves depositing the $V_2C$ material onto a thin 50 nm silicon oxide layer on a silicon substrate using drop casting and spin coating techniques to obtain individual flakes on the substrate. The sample is thoroughly cleaned with isopropyl alcohol and dried with nitrogen gas. Electrical contacts are fabricated using e-beam lithography patterning and e-beam metal deposition of Au/Ti (45 nm/5 nm) contacts on the $V_2C$ flakes **(Fig. 18a)**. A soft annealing process is performed at 90 °C for 5 minutes to coat the sample with an additional 100 nm layer of photoresist (PMMA) for passivation of the MXene flakes. The contact area is defined using e-beam lithography to create an open probing region for collecting photo carriers **(Fig. 18b** microscope image).

The fabricated device is electrically biased to observe the electrical dark current response, and its photocurrent response is measured under illumination with a 2 µm laser.

The obtained photocurrent response exhibits a significant change, which correlates with Fourier transform infrared (FTIR) spectroscopy measurements **(Fig. 18c)**. The device's maximum responsivity, $R_{ph}=I_{ph} - I_{dark}/P_{norm.laser}$, is determined to be approximately 2.65 A/W, where Pnorm.laser represents the normalized laser power incident on the device area. Normalizing the laser power helps accurately assess the material's responsivity by eliminating the effect of the variable sensing area.

To study the impact of channel length on photocurrent, several devices with varying channel lengths from 2-5 µm are fabricated. The results reveal a noticeable trend of increased photocurrent by a factor of $10^4$ when normalized with the sensing area along longer channel lengths. This provides insights into the scalability of device design for targeted sensing applications. Furthermore, these devices can be integrated into waveguides and photonic circuits for on-chip applications. Overall, our findings demonstrate the successful implementation of $V_2C$ as a promising material for mid-IR photodetection, highlighting its potential for future optoelectronic and integrated photonics applications.

**Modulator applications**

$Ti_3C_2T_x$ was employed as a saturable absorber in the schematic of the stretched-pulse mode-locked fiber laser shown in **Fig. 19a**. The laser pulses achieved a locked steady mode, as demonstrated in **Fig. 19(b,c)**, with a period of 50.37 ns and a temporal width of $10^4$ fs. All-optical modulators, which enable the control of one light's properties using another light without external electronic control, have garnered significant interest in communication and signal processing applications. In a peripheral study, a large nonlinear refractive index of $10^4$ was measured for $Ti_3C_2T_x$, facilitating the light-control-light system. **Fig. 19d** depicts a $Ti_3C_2T_x$-based all-optical modulator utilizing spatial cross-phase modulation. The nonlinear optical property of $Ti_3C_2T_x$ was activated by increasing the intensity of the red pump light, resulting in a diffraction pattern resembling the green probe light. The pump light had the ability to modify the ring numbers and diameters of the probe light. The manipulation of the probe light's phase change was achieved by modulating the power of the pump light, creating the MXene-based switcher **(Fig. 19e)**. A nonlinear four-wave mixing effect was employed to create an all-optical converter with exceptional nonlinear optical response in the telecommunication band. Through the integration of MXene, a conversion efficiency of 59 dB was achieved for transporting a 10 GHz signal **(Fig. 19f)**.

Furthermore, the high thermal conductivity and significant photothermal conversion efficiency of MXenes have been explored in the development of all-optical modulators. **Fig. 19g** illustrates the experimental setup for an all-optical modulator based on $Ti_3C_2T_x$. The fabricated modulator exhibited a modulation depth of >27 dB **(Fig. 19h)**, surpassing other Mach-Zehnder interferometer-based modulators. This was achieved by harnessing the photothermal effect. In a microfiber knot resonator (MKR) coated with $Ti_2CT_x$, the microfiber diameter was significantly reduced, indicating a faster response characteristic. The wavelength shift as a function of pump power is presented in **Fig. 19i**, with a measured pump-induced phase shift slope of 0.196 mW$^{-1}$. The rising and falling periods were estimated to be 306 and 301 s, respectively, as shown in **Fig. 19j**. These values are substantially faster compared to all-optical, all-fiber modulators utilizing other 2D materials such as graphene, SnS, WS2, and black phosphorus.

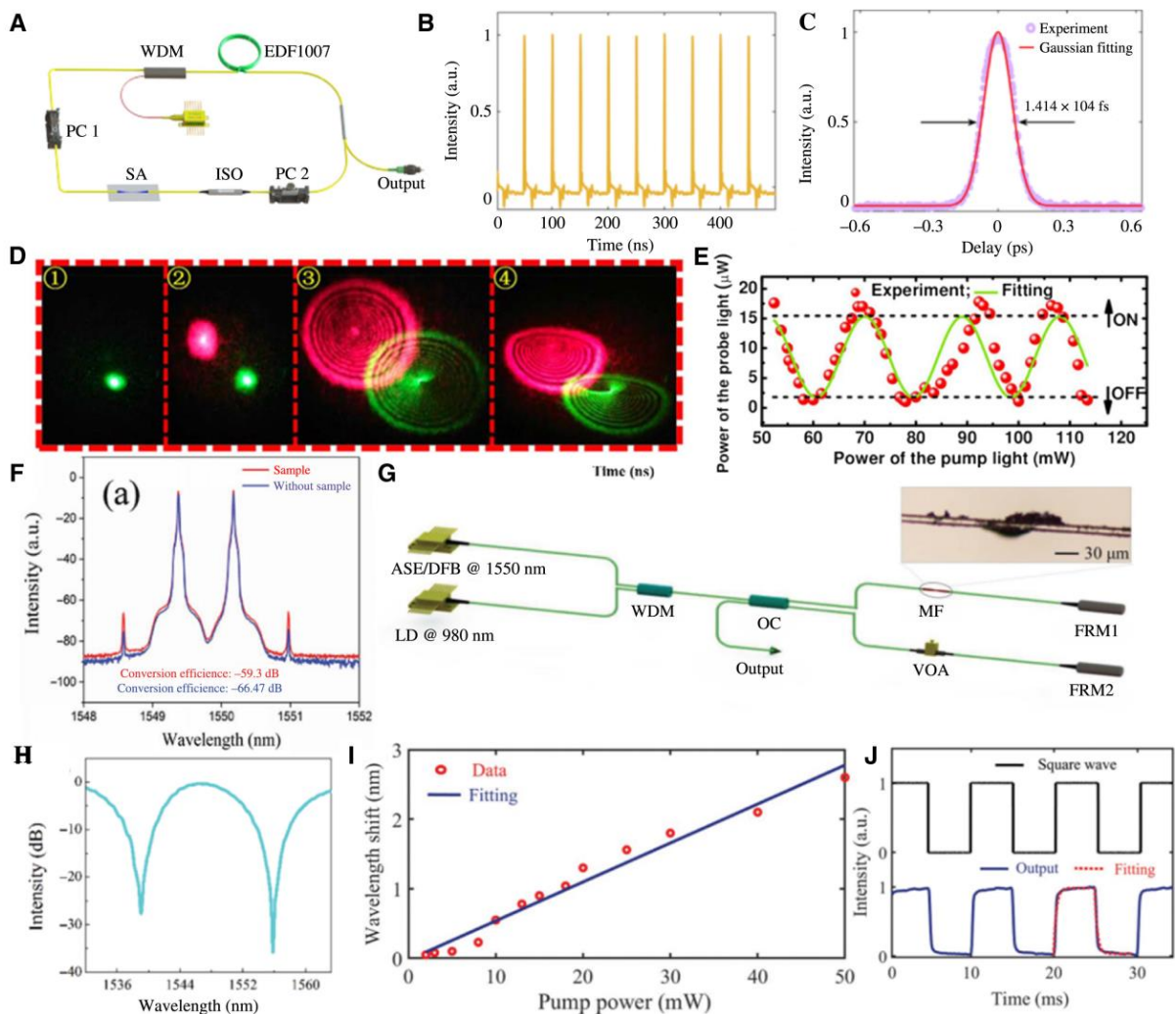

**Figure 19.** Modulators fabricated with MXenes material [60].

## Other photonic applications

Ti$_3$C$_2$Tx modified with soybean phospholipids exhibited excellent potential as a ceramic photothermal agent for cancer treatment, benefiting from its high photothermal conversion efficiency **(Fig. 20a)**. Moreover, the incorporation of Ti$_3$C$_2$Tx enhanced with soybean phospholipid demonstrated promising performance as a ceramic photothermal agent for cancer therapy, showcasing its high photothermal conversion efficiency **(Fig. 20b)**. Another application of Ti$_3$C$_2$Tx involves utilizing it in phase change materials made of polyoxyethylene, enabling the conversion and storage of solar energy. Remarkably, a high photothermal storage efficiency of up to 94.5% was achieved under solar irradiation. Figure 20C illustrates the schematic diagram of an integrated photothermal optical sensor utilizing Ti$_3$C$_2$Tx. By illuminating the sample with pumping light, Ti$_3$C$_2$Tx generated heat, inducing an additional phase shift in the Si waveguide. This resulted in a high control efficiency (0.19 mW$^{-1}$mm$^{-1}$). The use of MXene quantum dots (MQDs) has gained widespread popularity in various fields, including optoelectronics, bioimaging, and catalysis. To produce light ranging from blue to orange, MQDs were doped with sodium thiosulfate and ammonia water. **Fig. 20d** presents the design for manufacturing sodium thiosulfate- and ammonia water-doped MQDs. The photoluminescence spectrum in **Fig. 20e** depicts the greatest emission peaks of the doped MQDs in visible light emission. Furthermore, $Ti_3C_2T_x$ MQDs in ethanol exhibited intense white luminescence when illuminated by UV light, as shown in **Fig. 20f**. The strongest emission had a complete width at half maximum of 220 nm, with a central wavelength of 509 nm. MXenes have also demonstrated optical nonlinearities and the phenomenon of saturable absorption. At high optical intensities, MXenes induce optical loss in the light beam passing through them; however, the saturable absorption feature can be utilized to generate short pulses through passive mode locking [60].

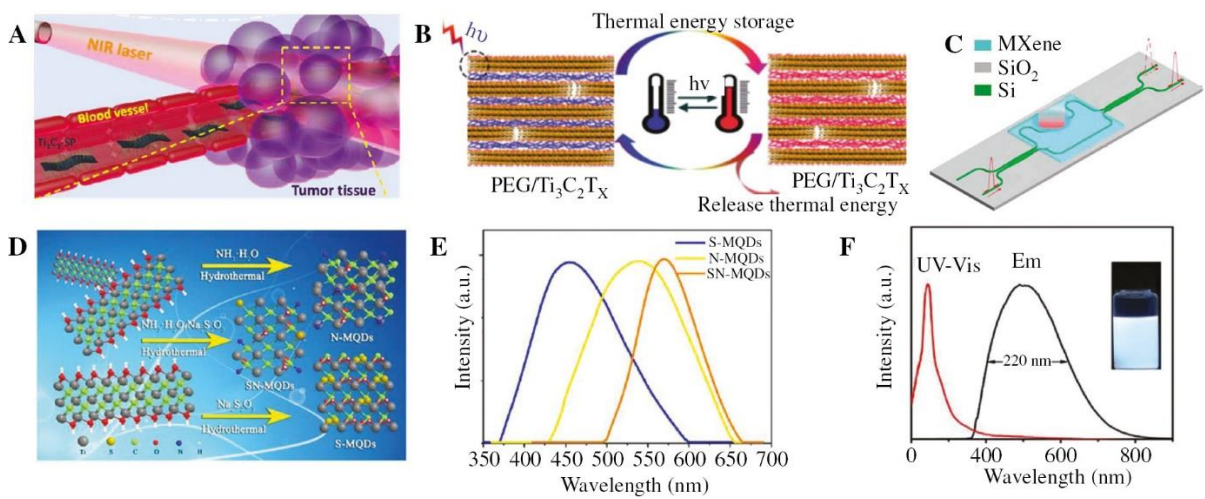

**Figure 20.** MXene applications for photothermal storage and QD LED [60].

## General Electronic Applications

MXenes have garnered significant attention as potential materials for electronic applications due to their unique properties, including metallic conductivity, hydrophilic surface, adjustable work function, and tunable interlayer spacing. These characteristics make MXene nanosheets suitable for the fabrication of thin layers in electronic devices. For instance, in the case of well-studied metallic MXene $Ti_3C_2Tx$, different work function values have been experimentally determined: 4.37 eV (measured by ultraviolet photoelectron spectroscopy (UPS) under vacuum), 4.60 eV (by photoelectron spectroscopy in air (PESA)), and 5.28 eV (by Kelvin Probe atomic force microscope (KPFM) under ambient air conditions). The variation in surface functional group composition and measurement environment/technique might account for the significant modulation of the work function (approximately 1 eV). Further comprehensive research is necessary to gain a deeper understanding of this behavior. Similarly, monolayer $MoS_2$ has been reported to exhibit different work function values under different conditions: 4.36 eV under air, 4.04 eV under ultrahigh vacuum, and 4.47 eV under oxygen ambient. These variations highlight the sensitivity of MXenes to their surroundings and emphasize the need for further investigations. The ability of MXenes to undergo charge transfer with adsorbates on their surface makes them attractive candidates for the development of innovative gas sensors based on Schottky diodes.

## Magnetic Devices Applications

While MXenes have been extensively studied for their optical and electrical properties, their magnetic characteristics have received less attention, resulting in a gap between theoretical hypotheses and experimental confirmations. Most of the predicted magnetic MXenes involve doping with magnetic transition metal elements such as Cr, V, Mn, Mo, Fe, Co, and Ni. However, due to limitations in material production processes, there are very few experimental reports on the magnetic properties of pure MXenes. A recent breakthrough in this area is the discovery of magnetism in reduced $Ti_3C_2$, which exhibited Pauli paramagnetism and a Curie-like increase in magnetism at low temperatures. This indicates that the magnetism originates from the 2D layers of $Ti_3C_2$ rather than the defects of $TiO_2$. The magnetic properties of MXenes hold promise for various applications, including magnetoresistance (MR) devices used in electronics for magnetic sensing and data storage. MR effects have been observed in Au-MXene-Au sandwich structures, with stable MR in the bias voltage range of 0.2 V to 1.0 V. The band gap and thickness of the MXene influence the MR effects. MXenes like $Mn_2CF_2/Ti_2CO_2/Mn_2CF_2$ are also being explored for magnetic tunnel junctions (MTJs) with high tunneling magnetoresistance ratios. Other interesting magnetic phenomena include the noncollinear 120 Y-type antiferromagnetic order in $Hf_2VC_2F_2$ MXene monolayers, which exhibit spin helical behavior, and the potential for multiferroicity where polarization and magnetism coexist. In graphene spintronics devices, the half-metallic monolayer $Cr_2C$ ferromagnet can be utilized to achieve high spin polarization. By incorporating $Cr_2C$ in a vertical graphene/$Cr_2C$ heterostructure, significant spin polarization of up to 74% has been observed. The strain-sensitive electrical structure and charge transfer in this heterostructure offer opportunities for strain-modulated spin filtering. While there is still much to explore in the field of MXene magnetism, these findings open up new possibilities for magnetic applications and spintronic devices **(Fig. 21)**.

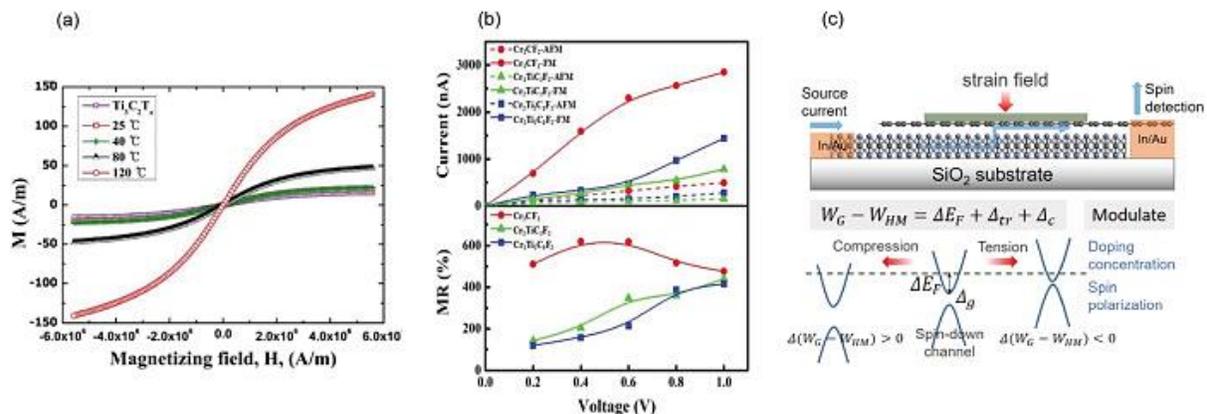

**Figure 21.** (a) Field-dependent magnetization curves (M–H) recorded at 300 K for the pristine $Ti_3C_2T_x$, and $Ti_3C_2T_x$ after reduction at different temperatures reveal a paramagnetic behavior. (c) Transport current (top

panel) and magnetoresistance value (bottom panel) versus the bias voltage for FM and AFM configurations in Au–Cr2CF2–Au sandwich structures.

## Conclusion and perspectives

Since their successful synthesis in 2011, MXenes have attracted significant research and attention, leading to advancements in their electronic and photonic applications, known as MXetronics. These two-dimensional transition metal carbides and nitrides possess unique properties, including a large surface area, diverse functionalization capabilities, desirable optical properties, and excellent biocompatibility. While MXenes have been extensively studied, their utilization in nanomedicine is still in the early stages of development. One promising future application is the use of MXenes in mid-infrared (mid-IR) photonics. There is ongoing research exploring the potential of MXenes as a source and detector in the mid-IR range, which will contribute to future advancements in this field. Additionally, experimental studies have demonstrated that $V_2CT_x$ MXene serves as an effective and stable electrocatalyst for nitrogen reduction reactions (NRR). These findings highlight the potential of $V_2C$ MXene in optoelectronic and integrated photonics applications, particularly for mid-IR photodetection.

## Acknowledgment

The Deanship of Scientific Research (DSR) at King Abdulaziz University (KAU), Jeddah, Saudi Arabia, funded this project under grant no. (KEP-PhD: 72-130-1443).

## Data Availability

The data that supports the findings of this study are available within the article.